\documentclass[aps,pra,twocolumn,showpacs,superscriptaddress,floatfix,10pt]{revtex4-1}
\usepackage{framed}
\usepackage{graphicx}
\usepackage{xfrac}
\usepackage{amsmath}
\usepackage{epsfig}
\usepackage{helvet}
\usepackage{amssymb}
\usepackage{soul}
\usepackage{xcolor}
\usepackage{framed}
\usepackage{hyperref}
\usepackage{amsthm}

\newcommand{\CC}{C\nolinebreak\hspace{-.05em}\raisebox{.4ex}{\tiny\bf +}\nolinebreak\hspace{-.10em}\raisebox{.4ex}{\tiny\bf +}}

\def\myfloor#1{\left\lfloor {#1} \right \rfloor}

\newtheoremstyle{mine}
  {\topsep}
  {\topsep}
  {\itshape}
  {0pt}
  {\bfseries}
  {:}
  {5pt plus 1pt minus 1pt}
  {}
\theoremstyle{mine}
\newtheorem*{conjecture}{Conjecture}

\begin{document}

\title{Holographic Cone of Average Entropies}

\author{Bart{\l}omiej Czech and Sirui Shuai}
\affiliation{Institute for Advanced Study, Tsinghua University, Beijing 100084, China}
\vskip 0.25cm

\begin{abstract}
\noindent
The holographic entropy cone identifies entanglement entropies of field theory regions, which are consistent with representing semiclassical spacetimes under gauge/gravity duality; it is currently known up to 5 regions. We point out that \emph{average} entropies of $p$-partite subsystems can be similarly analyzed for arbitrarily many regions. We conjecture that the holographic cone of average entropies is simplicial and specify all its bounding inequalities. Its extreme rays combine features of bipartite and perfect tensor entanglement, and correspond to stages of unitary evaporation of old black holes.
\end{abstract}

\maketitle
\textit{Introduction.---} One of the main approaches to quantum gravity is gauge/gravity duality or AdS/CFT correspondence \cite{adscft}. It posits that certain non-gravitational systems (conformal field theories, CFT) in $d$ dimensions provide a dual description of theories of gravity with anti-de Sitter (AdS) boundary conditions in $d+1$ dimensions. Perhaps the deepest insight into the fundamental nature of gravity ushered by the AdS/CFT correspondence is its intimate connection to quantum information theory. Succinctly put, gravitational spacetimes function like maps of quantum entanglement in the dual field theory \cite{marksessay, erepr}. On the other hand, most quantum systems do not have a dual gravitational description; only certain special systems are \emph{holographic} in that sense. This motivates a question at the intersection of information theory and gravity \cite{hec}: What necessary conditions must a quantum state satisfy if its quantum entanglement is to be holographically represented as a gravitational spacetime?

A technical intermediary between information theory and gravity is the Ryu-Takayanagi proposal \cite{rt1, rt2, hrt}, which is a holographic generalization of the Bekenstein-Hawking formula for black hole entropy \cite{bh1, bh2}. The proposal asserts that the von Neumann entropy of the reduced state of a CFT subregion is represented in the bulk AdS geometry as the area of the smallest extremal codimension-2 surface, which is homologous to the said boundary region. A fact in geometry is that areas of minimal surfaces homologous to fixed boundary regions automatically satisfy certain inequalities, for example \cite{mmiref}:
\begin{equation}
S_{AB} + S_{BC} + S_{CA} - S_{A} - S_{B} - S_{C} - S_{ABC} \geq 0 \label{monogamy}
\end{equation}
(Here $AB$ denotes the union of disjoint boundary regions $A$ and $B$.) The Ryu-Takayanagi proposal and the existence of a semiclassical bulk dual demand that the same inequalities be satisfied by entanglement entropies of CFT subregions. To be sure, states violating (\ref{monogamy}) and other {holographic inequalities} exist in quantum mechanics, but they cannot be consistently represented as areas of surfaces stipulated in the Ryu-Takayanagi proposal. 

Entropies of subregions, which are consistent with a semiclassical bulk interpretation, comprise the \emph{holographic entropy cone} \cite{hec}. An explanation of this terminology is that linear inequalities such as (\ref{monogamy}) are saturated on hyperplanes in entropy space---the vector space of hypothetical assignments of numbers (entropies) to regions. As a collection of hyperplanes, holographic inequalities bound a convex cone. The holographic entropy cone is currently known for up to five regions \cite{cuenca}. We also know one infinite family of inequalities for arbitrarily many regions \cite{hec} and several properties, which all holographic inequalities (known and unknown) must possess \cite{arrangement, kbasis, superbalance}. 

Yet progress in studies of the holographic entropy cone has been more quantitative than qualitative. With a few exceptions---inequality~(\ref{monogamy}) as a count of perfect 4-tensor entanglement \cite{mmipt}, and the infinite series of inequalities from \cite{hec} as discrete analogues of differential entropy \cite{diffent}---much of what we know about the holographic entropy cone is still waiting to be deciphered in heuristic terms. The difficulty in interpretation is partly due to a lack of data. If we knew the inequalities for more than five regions, we would presumably discern some patterns, which could then be examined for qualitative insights.

This paper makes a step in this direction. We identify a sector of the holographic entropy cone, which is exactly solvable for arbitrarily many CFT regions. Or so we believe; we conjecture the exact form of this sector, including all bounding inequalities and extreme rays, but do not give a proof. As a qualitative insight, our conjecture characterizes the most efficient purification consistent with a semiclassical bulk dual as coming from bipartite entanglement; see (\ref{lowerbound}) for a technical statement. Extreme rays achieve the most efficient purification on higher-partite subsystems but maximally violate it on lower-partite subsystems. This characterization of extreme rays implies that they describe stages of evaporation of old black holes. 

\textit{Holographic permutation invariants.---}
Our exactly solvable sector involves permutation invariants. The permutations in question simply relabel regions, for example $A \to B \to C \to A$. When we have $N$ named regions, the relevant permutation group is $S_{N+1}$. This is because every $N$-partite mixed state can be purified with the addition of an $(N+1)^{\rm st}$ system, whose label $O$ can also be switched with $A$, $B$, etc. We remark that switching a named region $A$ with the purifier $O$ generally rewrites inequalities in a nontrivial way. For example, for subadditivity---an inequality which holds true even outside holography, by virtue of quantum mechanics alone \cite{saref}---we have in the $(N=3)$-region context:
\begin{equation}
S_A \!+\! S_B \!-\! S_{AB} \geq 0 
\, \xrightarrow{A \to O = \overline{ABC}} \, 
S_{ABC} \!+\! S_{B} \!-\! S_{AC} \geq 0
\label{subadditivity}
\end{equation}
On the other hand, the monogamy of mutual information~(\ref{monogamy}) is $S_4$-invariant \cite{mmipt}.

Entropies of $N$-partite states contain $N/2$ or $(N+1)/2$ $S_{N+1}$-invariants, whichever is integer.  (We denote this number $\myfloor{(N+1)/2}$, where $\myfloor{\ldots}$ is the floor function.) The claim becomes obvious when we write the invariants explicitly. They are average entropies of $p$-component regions, henceforth denoted $S^p$, with $1 \leq p \leq \myfloor{(N+1)/2}$. For example, in the $(N=3)$-region context, we have:
\begin{align}
S^1 & = \tfrac{1}{4} \big(S_A + S_B + S_C + S_O\big) \label{defs1} \\
S^2 & = \tfrac{1}{6} \big(S_{AB} + S_{AC} + S_{AO} + S_{BC} + S_{BO} + S_{CO} \big) \label{defs2}
\end{align}
Note that, if we avoid using the purifier explicitly in formulas, $S^p$ contains entropies of complementary $(N+1-p)$-partite regions. For example, (\ref{defs1}-\ref{defs2}) are equivalent to:
\begin{align}
S^1 & = \tfrac{1}{4} \big(S_A + S_B + S_C + S_{ABC}\big) \label{defs1nO} \\
S^2 & = \tfrac{1}{3} \big(S_{AB} + S_{AC} + S_{BC}\big) \label{defs2nO}
\end{align}
The equivalence between $p$-partite and $(N+1-p)$-partite regions' entropies explains why the count of permutation invariants only goes up to $\myfloor{(N+1)/2}$. 

\textit{Inequalities on average entropies.---}
A typical holographic inequality concerns more than just permutation invariants. But it is easy to convert it to a statement about averages: we simply replace every $p$- and $(N+1-p)$-component term in the inequality with $S^p$. Thus, for subadditivity~(\ref{subadditivity}) and monogamy~(\ref{monogamy}) we have:
\begin{align}
2S^1 - S^2 & \geq 0 \label{invsa} \\
- 4S^1 + 3S^2 & \geq 0 \label{invmono}
\end{align}
To confirm the validity of this substitution, observe that if an entropy vector obeys an inequality then so do all of its $S_{N+1}$-images. Replacing a $p$-component entropy with $S^p$ applies the original inequality to the average of all permutation images.

\textit{Holographic cone of average entropies.---}
While all holographic inequalities---known and unknown---carve out the full holographic entropy cone, the permutation-invariant inequalities carve out a smaller cone: the cone of average entropies. This cone is a projection of the full holographic entropy cone to the subspace of entropy space spanned by $S^p$. Its bounding inequalities, such as (\ref{invsa}) and (\ref{invmono}), constrain the ratios of average $p$-partite entropies. For $N$ named regions, the cone of averages lives in an $\myfloor{(N+1)/2}$-dimensional linear space. 

\begin{conjecture}
The holographic cone of average entropies, for any number $N$ of regions, is as follows:
\end{conjecture}
\vspace*{-8pt}
\begin{itemize}
\item The cone is simplicial. That is, it has $\myfloor{(N+1)/2}$ bounding facets (maximally tight inequalities) and $\myfloor{(N+1)/2}$ extreme rays---loci where $\myfloor{(N-1)/2}$ inequalities are saturated. 
\item The extreme rays are obtained by minimal cuts on weighted graphs shown in Figure~1. We call them \emph{flowers} with one \emph{stem} of weight $w$ and $N$ \emph{petals} of weight 1. The weights $w$, which realize extreme rays, are all odd (or even) positive integers up to and including $N$. 
\item The bounding facets of the cone are generated by subadditivity ($F_1 \equiv S^1 - S^2/2 \geq 0$) and novel inequalities indexed by $2 \leq p \leq \myfloor{(N+1)/2}$:
\begin{equation}
F_p \equiv \frac{2 S^p}{p} - \frac{S^{p-1}}{p-1} - \frac{S^{p+1}}{p+1} \geq 0
\label{newpineq} 
\end{equation}
($F_2 \geq 0$ symmetrizes monogamy (\ref{monogamy}) while $F_3 \geq 0$ symmetrizes known inequalities from the $N=5$ holographic entropy cone \cite{cuenca}.) Supplemental Material \cite{sm} verifies that the conjectured extreme rays and facets are complete and consistent.
\end{itemize}

\begin{figure}[!t]
\label{flowersgraph}
\begin{center}
\includegraphics[width=0.6\columnwidth]{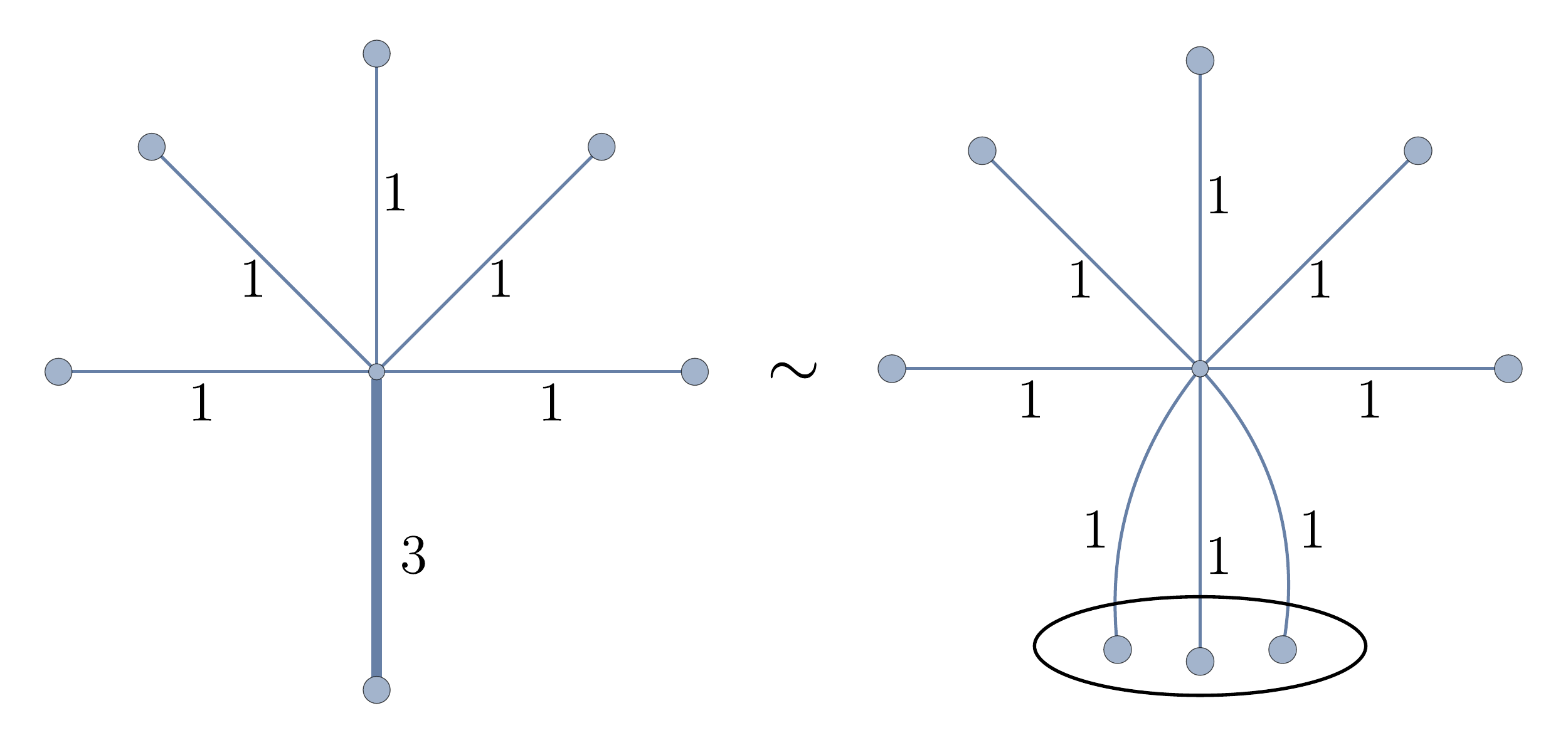}
\hspace*{0.05\columnwidth}
\includegraphics[width=0.3\columnwidth]{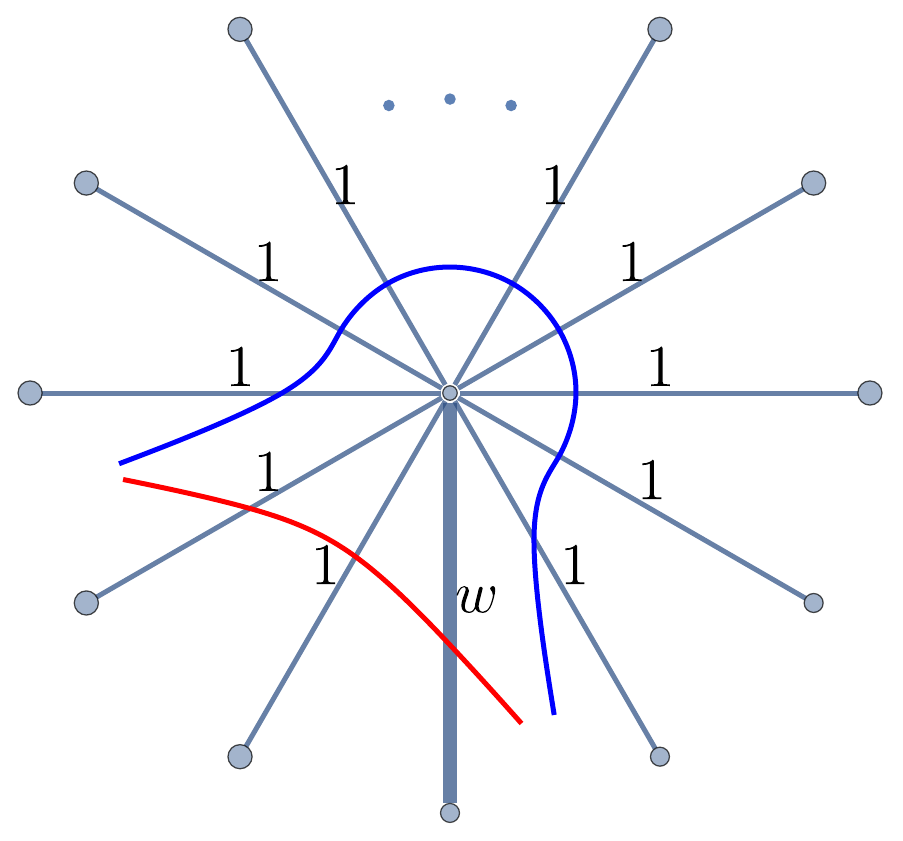}
\caption{A flower graph with $N$ `petals' of weight 1 and one `stem' of weight $w$ (here $w=3$) can be obtained from a perfect tensor graph with $N+w$ legs by bundling up $w$ legs. The entropy of each $p$-partite region is the smaller of two cuts, which sever $p$ or $N+1-p$ legs. Averages of these cuts realize extreme vectors of the holographic cone of average entropies.}
\end{center}
\end{figure}

Inequality~(\ref{newpineq}) requires a minor clarification. When $p = \myfloor{(N+1)/2}$, $(p+1)$-composite regions are related by $S_{N+1}$ permutations to $(N-p)$-composite regions. (We saw this in equations~\ref{defs1nO}, \ref{defs2nO}). We may state this fact as an equivalence between quantities $S^p$:
\begin{align}
S^{(N+2)/2} & \equiv S^{N/2} \quad && (N~{\rm even}) \\
S^{(N+3)/2} & \equiv S^{(N-1)/2} \quad && (N~{\rm odd})
\end{align}
Thus, in the special case of $p = \myfloor{(N+1)/2}$, inequality~(\ref{newpineq}) takes superficially modified forms:
\begin{align}
(N+3) (N-1)\, S^{(N+1)/2} \,-\, (N+1)^2 S^{(N-1)/2} & \geq 0 \,\,\,\,
\nonumber \\
(N~{\rm odd}) \quad & \label{specialodd} \\
(N+4) (N-2)\, S^{N/2} \,-\, N(N+2)\, S^{(N-2)/2} & \geq 0 \,\,\,\,
\nonumber \\
(N~{\rm even}) \quad & \label{specialeven}
\end{align}

In Figure~2 and Supplemental Material \cite{sm} we show the cones of average entropies up to $N=6$ named regions. They conform with our conjecture. We also display known inequalities and rays pertaining to the $N=7$ cone of averages. There is a small region in the space of averages not excluded by previously known inequalities, for which we have been unable to find a consistent assignment of entropies. (Following \cite{hec}, every entropy vector in the holographic entropy cone {should} be realizable by minimal cuts on a weighted graph.) Inequality~(\ref{specialodd}) with $N=7$ eliminates that questionable region. These remarks establish that our conjecture is consistent with known facts. 

\begin{figure}[!t]
\label{n5fig}
\begin{center}
\includegraphics[width=0.45\columnwidth]{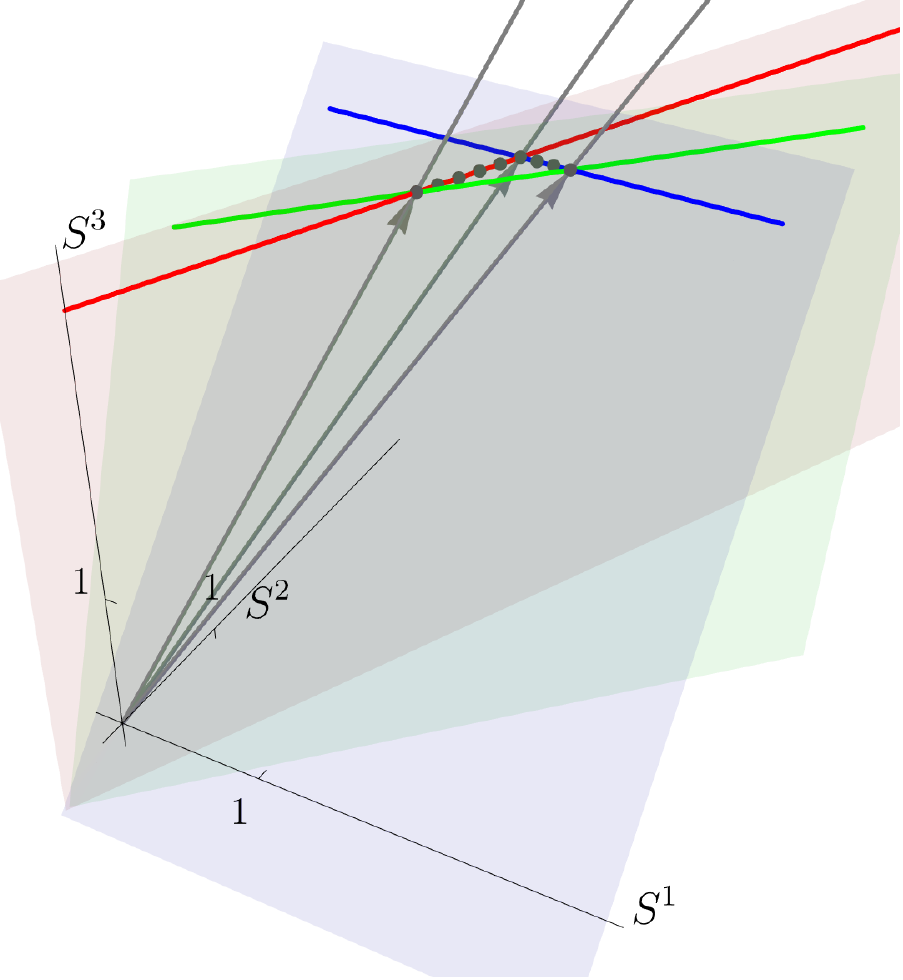}
\hspace*{0.05\columnwidth}
\includegraphics[width=0.45\columnwidth]{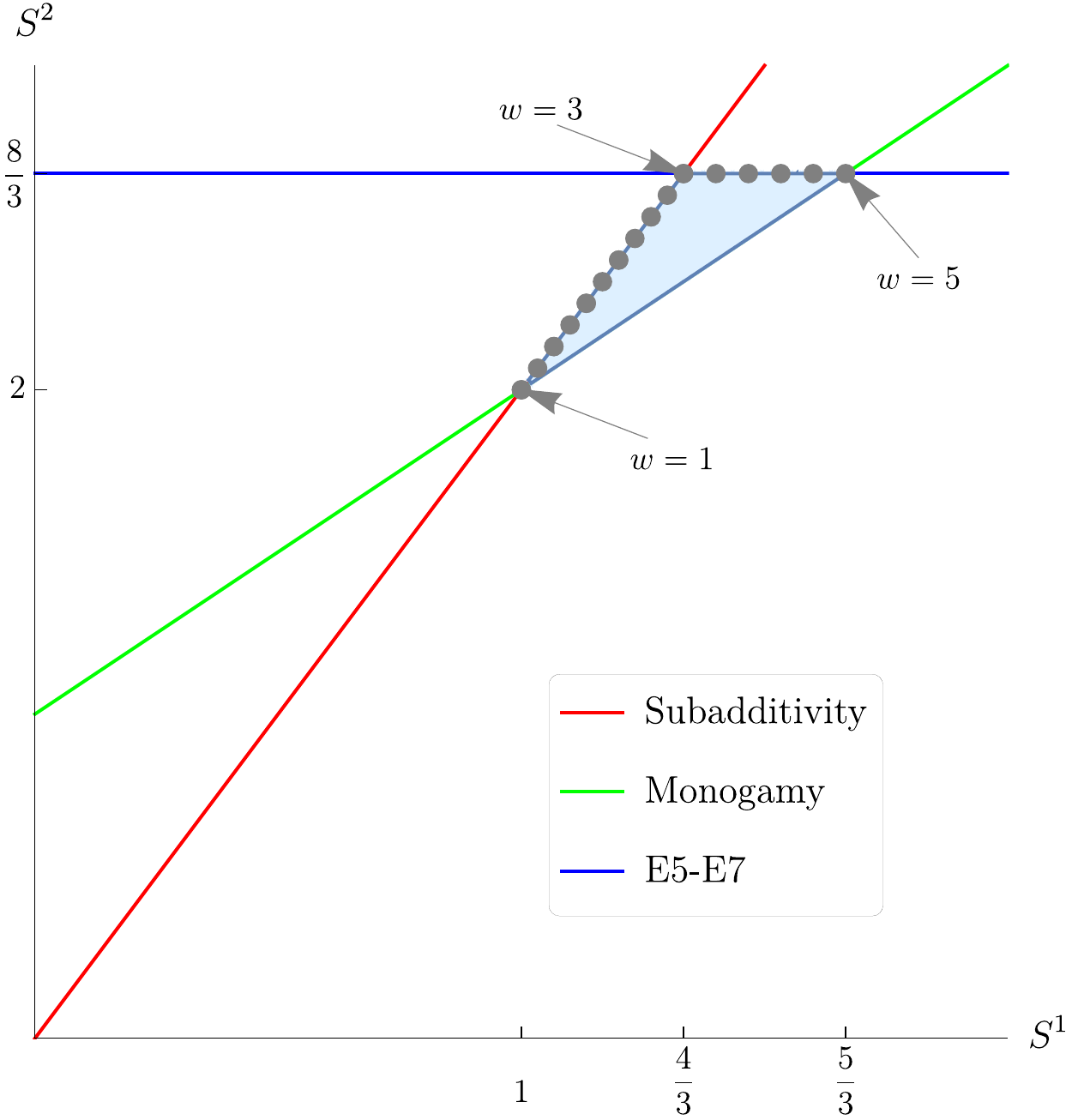}
\caption{The $N=5$ holographic cone of average entropies is bound by three planes in three dimensions, which originate from subadditivity, monogamy, and inequalities E5-E7 from \cite{cuenca}. Cuts through flower graphs with varying weight $w$ (shown as a series of dots) trace out the simplicial shape of the cone's cross-section. The right panel is the $S^3=3$ plane.}
\end{center}
\end{figure}

\textit{The conjecture constrains efficiency of purification.---} 
To elucidate the meaning of our conjecture, we consider conditional entropy $S(Y|X) = S_{YX} - S_X$. It characterizes region $Y$'s contribution to purifying $X$. Conditional entropy equals $S_Y$---the maximum allowed by subadditivity--when $Y$ does not purify $X$ at all. To minimize $S(Y|X)$ is to find a $Y$, which best purifies $X$.  

The \emph{average conditional entropy} of one region, conditioned on a $p$-partite system, is $S^{p+1} - S^p$. This quantity captures how much, on average, adjoining a single region purifies a $p$-partite system. Using strong subadditivity \cite{ssaref}, which holds generally in quantum mechanics, one can prove that it can never exceed $S^p / p$, or equivalently:
\begin{equation}
\frac{S^{p+1}}{S^{p}} \leq \frac{p+1}{p}
\label{upperbound}
\end{equation}
In Supplemental Material \cite{sm} we prove (\ref{upperbound}) and show that it saturates if and only if all regions $X$ and $Y$, which together cover $p+1$ or fewer basic constituents, have zero mutual information (do not help to purify one another).

Now observe that inequalities~(\ref{newpineq}) also imply (\ref{upperbound}):
\begin{equation}
\frac{S^{p}}{p} - \frac{S^{p+1}}{p+1} = \sum_{p'=1}^p F_{p'}
\label{upperrewrite}
\end{equation}
By itself, this fact does not characterize our conjecture because---we stress---(\ref{upperbound}) follows from strong subadditivity alone. It does, however, set an illuminating counterpoint to the \emph{holographic lower bound} on $S^{p+1} - S^p$:
\begin{equation}
\frac{S^{p+1}}{S^{p}} \geq \frac{p+1}{p} \cdot \frac{N-p}{N-p+1}
\label{lowerbound}
\end{equation}
Bound~(\ref{lowerbound}) also follows from inequalities~(\ref{newpineq}):
\begin{align}
& \frac{S^{p+1}}{(p+1)(N-p)} - \frac{S^{p}}{p(N-p+1)} \qquad\qquad\qquad (N~{\rm odd}) \nonumber \\
& \propto \frac{N+3}{4} F_{\myfloor{(N+1)/2}} + \sum_{p' = p+1}^{\myfloor{(N-1)/2}} (N+1-p') F_{p'}
\label{lowerrewrite}
\end{align}
(For even $N$ the coefficient of $F_{N/2}$ is $(N+2)/2$.) But unlike~(\ref{upperbound}), (\ref{lowerbound}) cannot be derived from previously known entropic inequalities, general or holographic. 

Owing to rewritings (\ref{upperrewrite}) and (\ref{lowerrewrite}), bounds (\ref{upperbound}) and (\ref{lowerbound}) enjoy many parallels. Whereas saturating (\ref{upperbound}) at $p$ implies saturation of the same bound for all $p' < p$, {saturating (\ref{lowerbound}) at $p$ implies saturation of the same bound for all $p' > p$.} (The latter statement holds if and only if our conjecture does.) Whereas saturating (\ref{upperbound}) describes when adjoining one extra region is least helpful in purifying $p$-partite systems, saturating (\ref{lowerbound}) describes when it is most helpful. In the form (\ref{lowerbound}), our conjecture describes the most efficient rate of purifying $p$-partite systems (by the addition of one constituent), which is consistent with a semiclassical bulk dual.

%

\begin{figure}[!t]
\label{flowersgraph}
\begin{center}
\includegraphics[width=0.22\columnwidth]{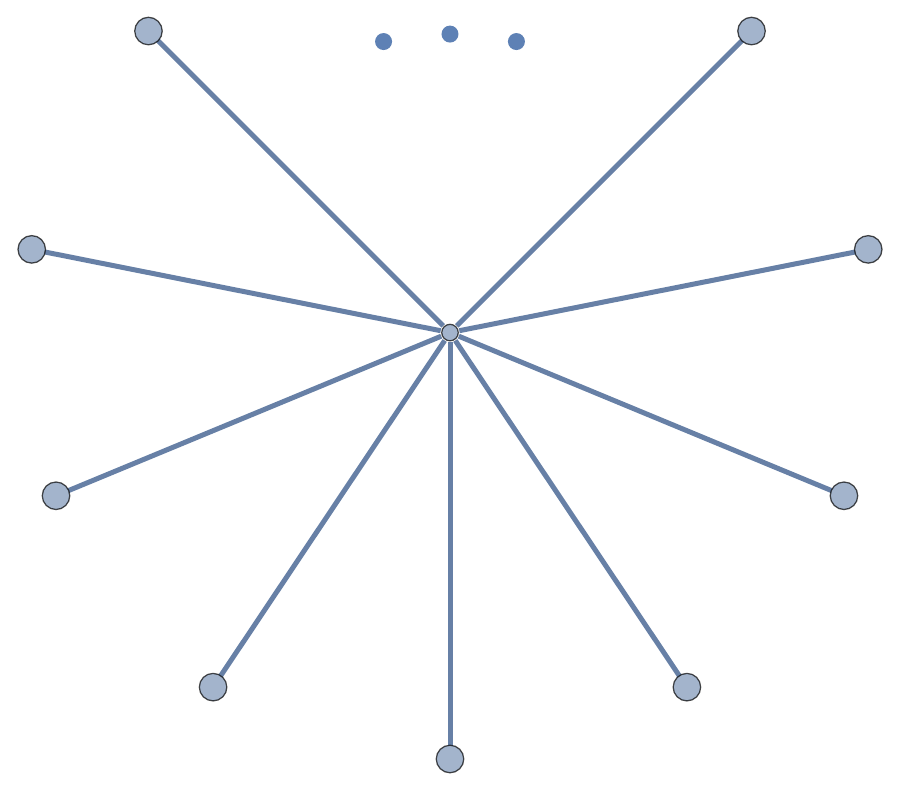}
\hspace*{0.15\columnwidth}
\includegraphics[width=0.6\columnwidth]{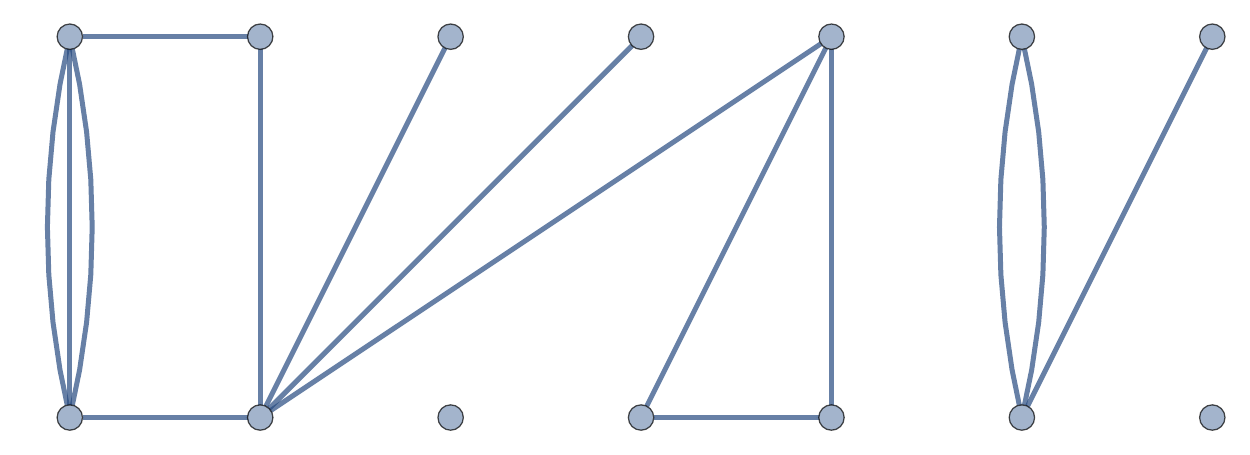}
\caption{Perfect tensor (PT, left) and bipartite (EPR, right) entanglement; eqs.~(\ref{defpt}-\ref{defepr}). Here all edges have equal weight.}
\end{center}
\end{figure}

\textit{Patterns of maximal and minimal purification.---} The fact that bounds (\ref{upperbound}) and (\ref{lowerbound}) can only be saturated on entire ranges of $p$ is significant. We shall see momentarily that it is responsible for the simplicial character of the holographic cone of average entropies. As a preliminary, we define two special patterns of entanglement, where (\ref{upperbound}) and (\ref{lowerbound}) are saturated for all $1 \leq p \leq \myfloor{(N+1)/2}$:
\begin{align}
{\rm PT:}& \quad S^p \propto p & {\rm for}~p \leq \myfloor{(N+1)/2} & \,\, \label{defpt}\\
{\rm EPR:}& \quad S^p \propto p(N+1-p)  & \textrm{for all}~p & \,\,\label{defepr}
\end{align}
Note that both PT and EPR saturate inequalities~(\ref{newpineq}). Our conjecture thus distinguishes (\ref{defpt}, \ref{defepr}) as two extreme entanglement patterns, which are uniform over $p$.

The label `PT' stands for perfect tensor entanglement. It is known to play an important role in holography; see for example \cite{happy, random, mmipt}. In the holographic entropy cone, even-membered perfect tensor states are extreme rays and form a complete basis for entropy space \cite{kbasis}. We illustrate them as weighted graphs in the left panel of Figure~3. PT is the pattern of least efficient purification.

The label `EPR' stands for bipartite entanglement. The acronym follows common nomenclature, which refers to \cite{eprref}. Here is why $S^p \propto p(N+1-p)$ describes EPR-like entanglement. A handy way to track bipartite entanglement is by drawing lines that connect pairs of regions; see the right panel of Figure~3. Such lines have been studied extensively in the literature following \cite{bitthreads} and are known as bit threads. Suppose the average number of bit threads between any two distinct regions is $K$. The EPR-like assumption on the entanglement structure means that a $p$-partite entropy equals the number of threads, which connect the $p$ constituents of the region with the $N+1-p$ constituents of the complement. For the average, this gives $S^p = K p (N+1-p)$. 

Armed with definition~(\ref{defepr}), we restate our conjecture in one final way, advertised in the introduction:

\begin{conjecture}
The most efficient rate of purifying mixed multi-partite states, which is consistent with a semiclassical bulk dual, is set by EPR-like entanglement. 
\nonumber
\end{conjecture}

\begin{figure}[!t]
\label{flowersgraph}
\begin{center}
\includegraphics[width=0.42\columnwidth]{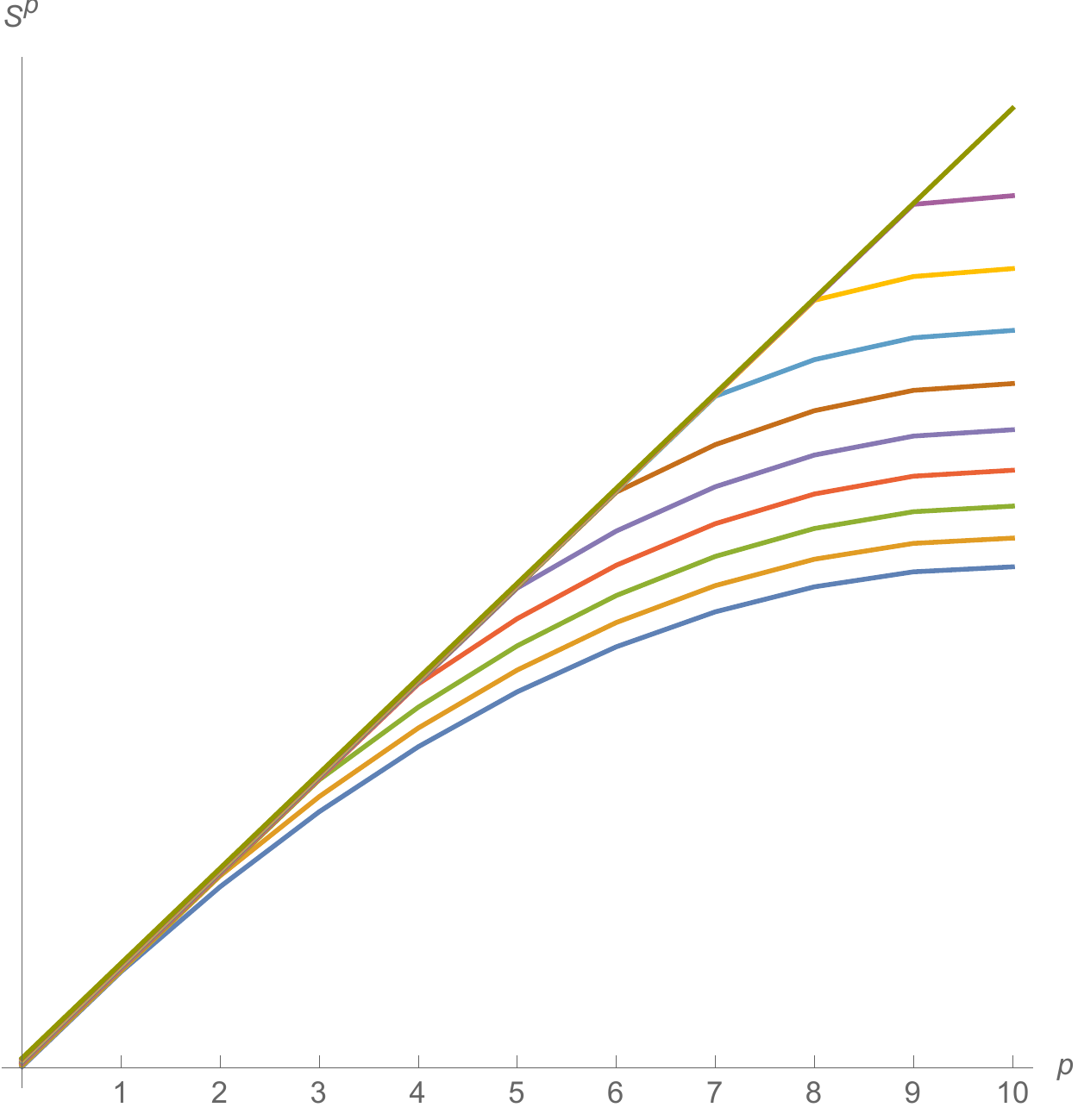}
\hspace*{0.08\columnwidth}
\includegraphics[width=0.42\columnwidth]{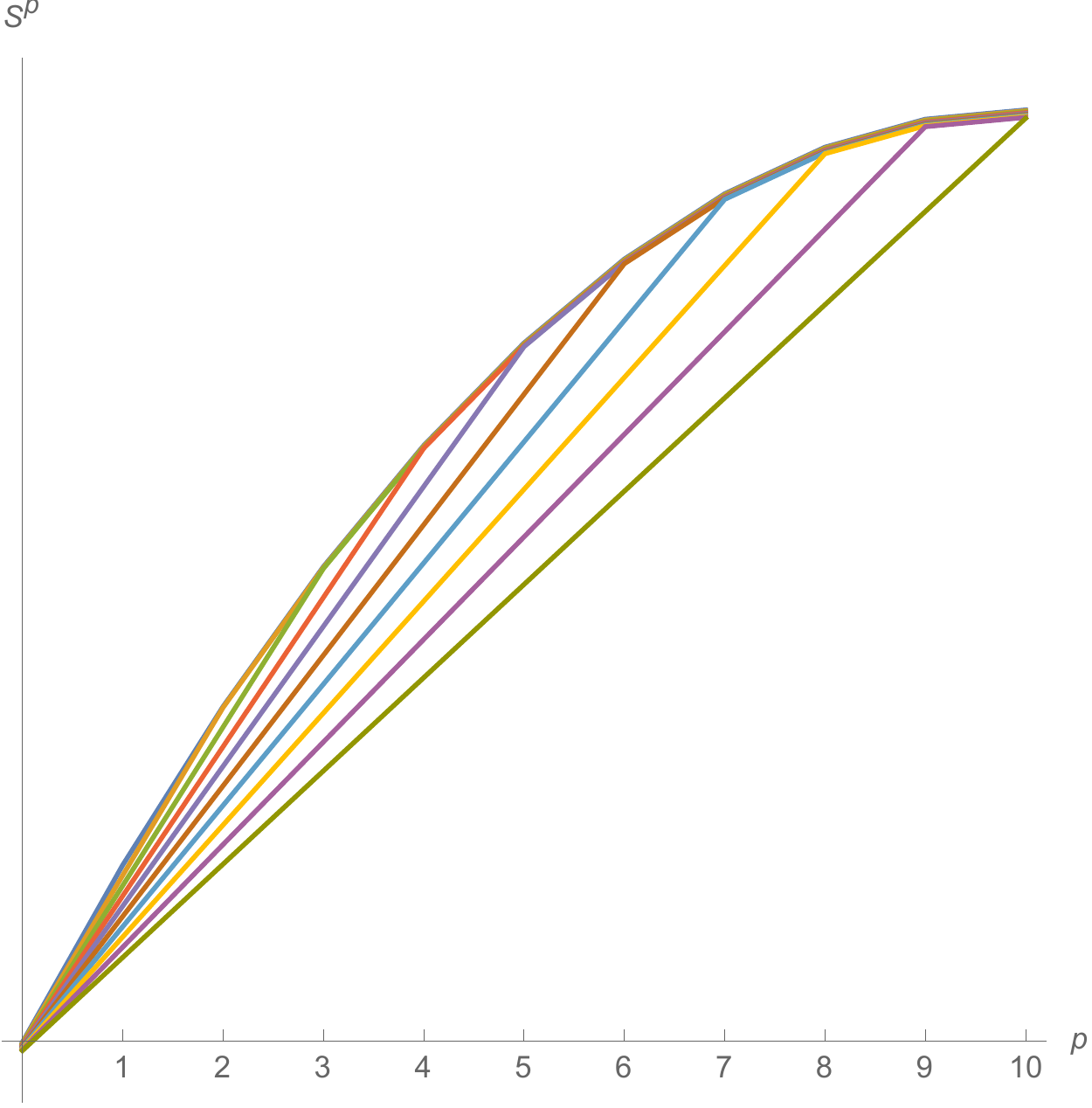}
\caption{Flower graphs are characterized by a threshold value $\tilde{p}$, which separates the linear and quadratic dependence of $S^p$ in equation~(\ref{svector}). Because the overall normalization of $S^p$ is tunable, we display $S^p$ of flower graphs normalized to fix $S^1$ (left) and $S^{\myfloor{(N+1)/2}}$ (right), taking $N = 19$.}
\end{center}
\end{figure}

\textit{Structure of the cone, reexamined---} A transparent way of presenting the average entropies $S^p$ is to plot them over the $p$-axis. We display such plots in Figure~4. In a general state with a semiclassical bulk dual, the slope of this plot at any $p$ is bounded above and below by (\ref{upperbound}) and (\ref{lowerbound}). Extreme plots are those, which saturate either the lower or the upper bound at each $p$. Because extremizing (\ref{upperbound}) always happens over a range $1 \leq p \leq \tilde{p}$ and extremizing (\ref{lowerbound}) always happens over a range $\tilde{p} \leq p \leq \myfloor{(N+1)/2}$, such {extreme plots} are uniquely specified by a single threshold parameter $\tilde{p}$:
\begin{equation}
S^p \propto \left\{\begin{array}{lp{1cm}l} p(N+1-\tilde{p}) && p \leq \tilde{p} \\
                                                                p(N+1-p) && p \geq \tilde{p} \end{array}\right.
\label{svector}
\end{equation}
In other words, an extreme vector of the cone of average entropies must be PT-like for $p \leq \tilde{p}$ and EPR-like for $p \geq \tilde{p}$. We verify in Supplemental Material \cite{sm} that this is exactly the dependence of $S^p$ in the flower graphs of Figure~1, with the threshold related to stem weight $w$ via $\tilde{p} = (N+2-w)/2$. The holographic cone of average entropies is simplificial because there are precisely $\myfloor{(N+1)/2}$ possible cross-overs $\tilde{p}$ to separate the PT- and EPR-like behaviors. The actual EPR pair (respectively perfect tensor) is recovered by setting $\tilde{p} = 1$ (respectively $\tilde{p} = \myfloor{(N+1)/2}$). 

This description of flower graphs adds a heuristic justification to our conjecture. We claim that (\ref{lowerbound}) identifies the most efficient way of purifying $p$-partite states. We know that (\ref{lowerbound}) is saturated by EPR pairs. We can still saturate (\ref{lowerbound}) yet depart from EPR-like entanglement (\ref{defepr}) by using $p$-local operations to suppress any $p$-local mutual information. We posit that such $p$-local operations do not help in purifying $p$-partite systems. 

\begin{figure}[!t]
\label{flowersgraph}
\begin{center}
\includegraphics[height=0.18\columnwidth]{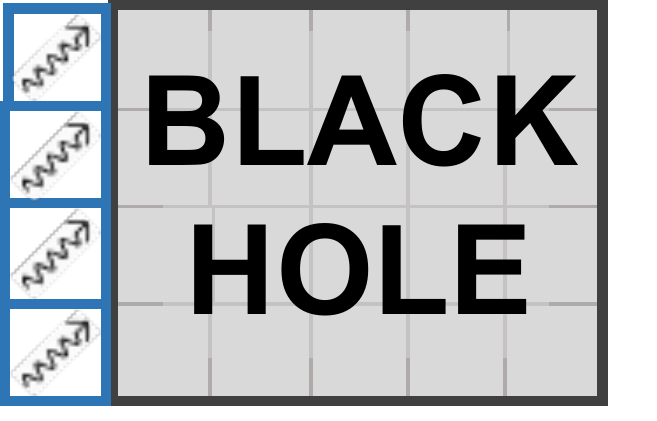}
\hspace*{0.06\columnwidth}
\includegraphics[height=0.18\columnwidth]{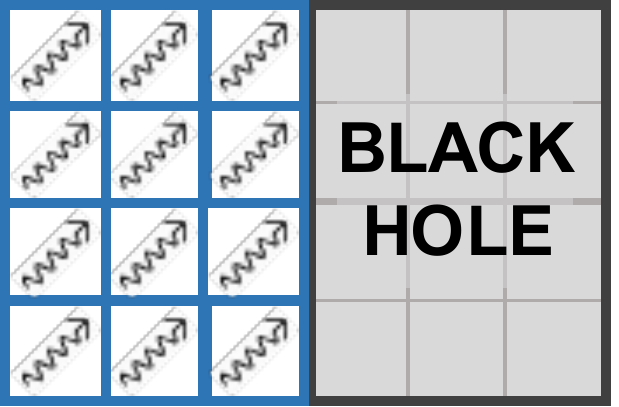}
\hspace*{0.06\columnwidth}
\includegraphics[height=0.18\columnwidth]{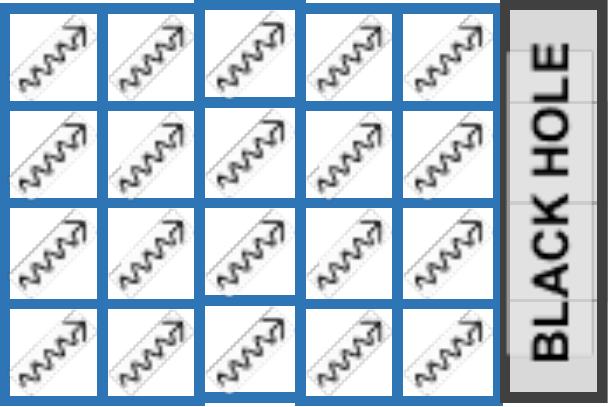}
\\
\includegraphics[width=0.27\columnwidth]{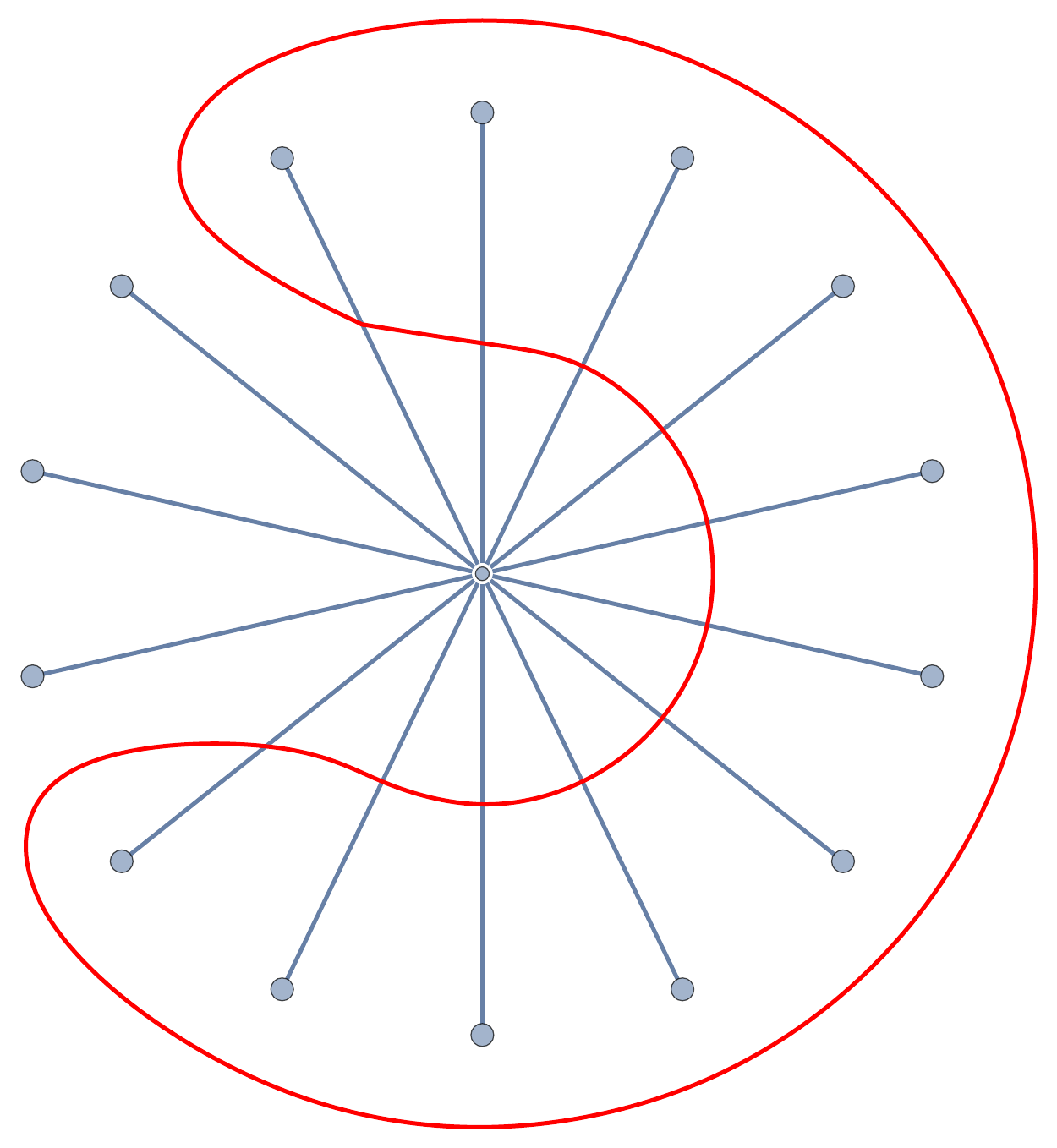}
\hspace*{0.05\columnwidth}
\includegraphics[width=0.27\columnwidth]{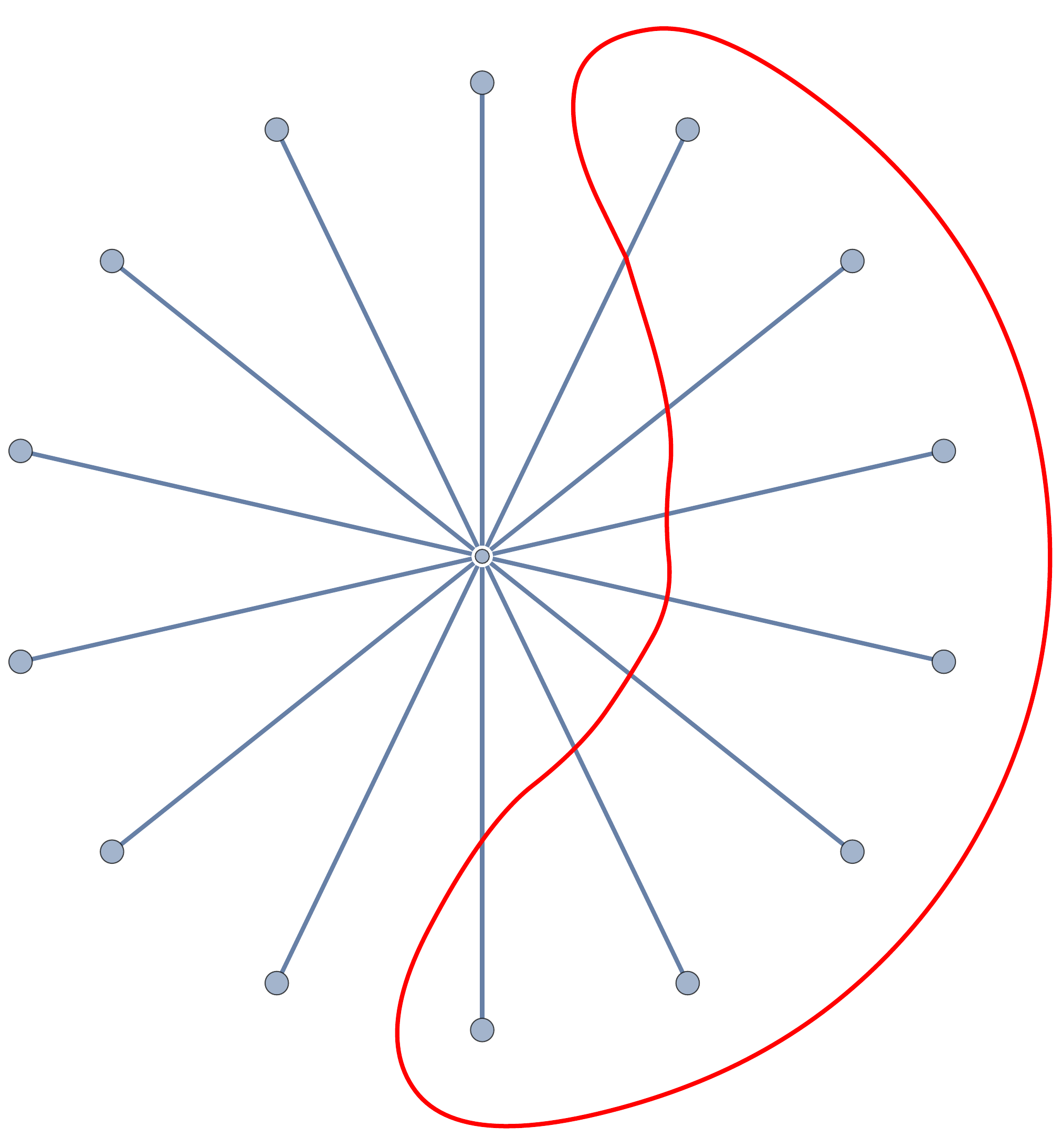}
\hspace*{0.05\columnwidth}
\includegraphics[width=0.27\columnwidth]{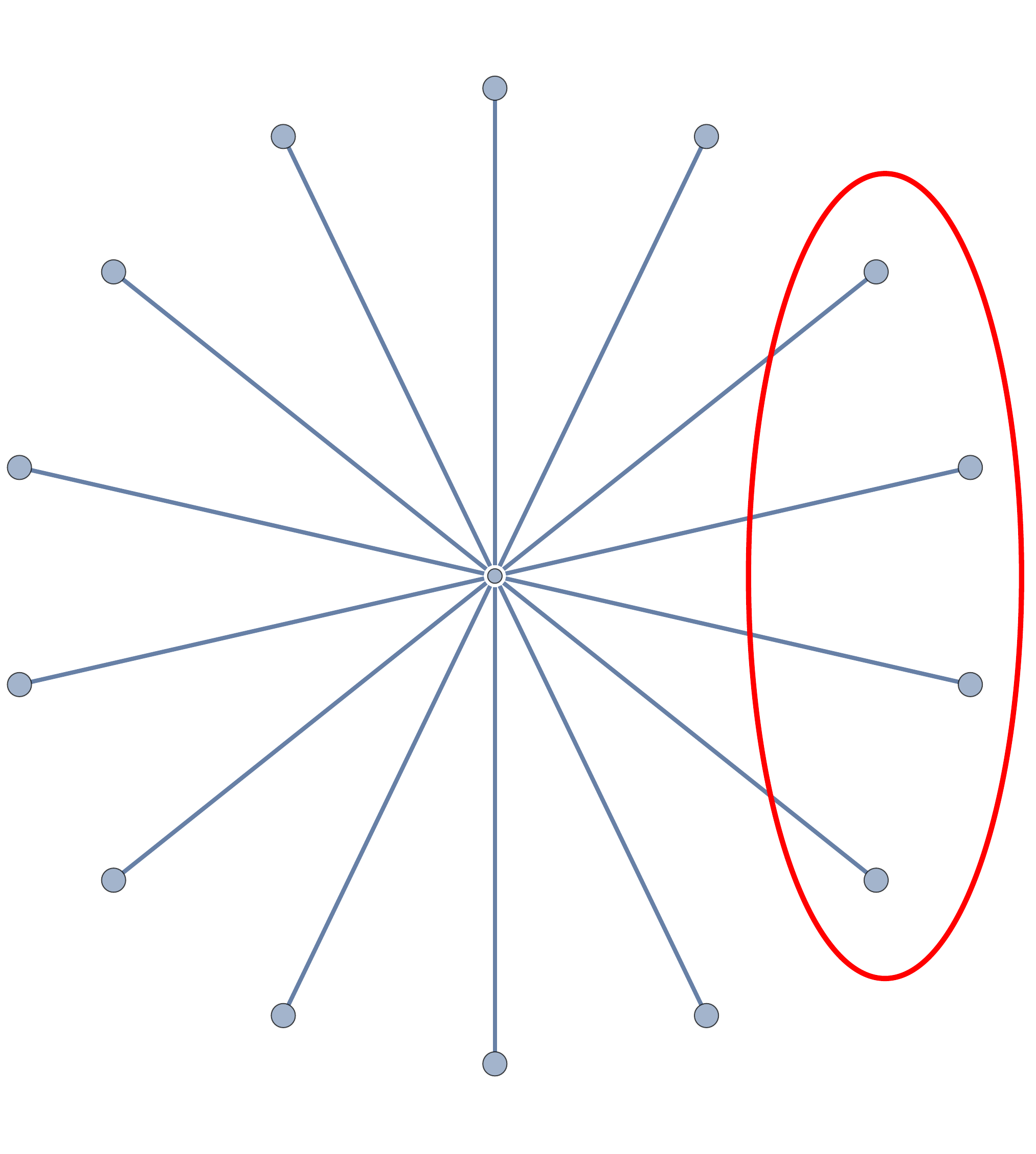}
\caption{Page \cite{page1} discusses black hole evaporation as a pure state in a box, divided into two parts: the black hole (whose size decreases in time) and radiation (which is complementary so its size increases in time). By Page's theorem \cite{pagethm}, the entropy of radiation in a random such state is the smaller of the two sizes, which gives the Page curve. A sliding division of a random multi-component state effectively constructs flower graphs. In this figure, time flows from left to right.}
\end{center}
\end{figure}

\textit{Cone of average entropies vs. black hole evaporation.---}
We would like to situate the holographic cone of average entropies in a broader physical context. As the extreme rays are organized by $p$-dependence of mutual information, we may anticipate that the cone will be relevant to information scrambling \cite{scrambling} (dilution of mutual information) and to radial depth (because $p$ captures a degree of nonlocality in the CFT). These heuristics suggest inspecting the cone from the viewpoint of black hole physics. And indeed, it turns out that the extreme rays of our cone correspond precisely to stages of unitary evaporation of old black holes, as described by Page \cite{page1, page2} and more recently in the islands proposal \cite{penington, princetonucsb, princeton}.

To explain this connection, we need one fact about flower graphs: that they can be obtained by grouping together $w$ constituents of a perfect tensor state on $N+w$ parties \cite{thankxlq}. This is illustrated in the middle panel of Figure~1. We argue that the same type of grouping is an essential aspect of black hole evaporation {\`a} la Page; see Figure~5. 

The argument in \cite{page1} considers a globally pure state of the combined black hole (BH) plus radiation (R) system. By unitarity, the global state remains pure at all times. Because the black hole evaporates, BH makes up a decreasing proportion of the total Hilbert space. Assuming the global state is nearly random, Page's theorem \cite{pagethm} tells us that the entropy of the BH system is the size of BH or the size of R, whichever is smaller. This yields the famous Page curve.  Now, a random state in a large Hilbert space has nearly perfect tensor entanglement \cite{random}. In isolating BH from a random state of the total system, Page effectively groups together $A_{\rm BH} \equiv w$ constituents of a perfect tensor, where $A_{\rm BH}$ is the horizon area. This ends up constructing flower graphs.

\textit{Comments.---} The petals of the flower are small constituents of a global pure state---that is, particles of Hawking radiation. The idea of isolating Hawking radiation from the boundary CFT is what propelled the islands proposal of \cite{penington}; it is also what makes the connection with flower graphs explicit. A transparent way to compare flower graphs with stages of black hole evaporation is to consult the geometrized model of the latter formulated in \cite{octopi}. That model involves `octopus' diagrams, which are identical to our flower graphs. 

Both here and in \cite{page1, octopi} the Hawking particles are assumed uncorrelated and identical. The `uncorrelated' assumption is motivated by the equivalence principle---`no drama' in the language of \cite{firewalls}. But the `identical' assumption can be removed. In Figure~5 the black hole arises from bundling together many constituents of a parent perfect tensor state. We may also consider bundling up Hawking particles into bins of distinct sizes. This process, which further breaks permutation symmetry, can yield extreme rays of the full holographic entropy cone.

Observe that flowers with $w > N$ have the same entropies as the flower with $w = N$. The $A_{\rm BH} > N$ regime is when the horizon area exceeds the total Hawking radiation; this is a black hole before Page time. Thus, flower graphs describe stages of evaporation of an \emph{old} black hole. 

Finally, we ask: is it surprising that the holographic entropy cone `knows about' black hole evaporation? We think not. Its extreme vectors are supposed to be marginally consistent with a semiclassical bulk interpretation. If not black holes, what other spacetimes might come closest to a breakdown of semiclassical gravity?

\textit{Why conjecture and not prove?---} Reference~\cite{hec} identified a mechanical technique for proving candidate inequalities, which involves constructing so-called contraction maps. We have used two implementations of a greedy algorithm that searches for contractions \cite{bogdanscontractor, michaelscontractor}. Both of them overloaded our standard Mac computers before proving or disproving the simplest inequality we conjecture: inequality~(\ref{newpineq}) with $p=4$. It may be more realistic to identify valid non-symmetric inequalities, which reduce to (\ref{newpineq}) under symmetrization, and try to prove them by contraction. For example, the $p=4$ instance of (\ref{newpineq}) predicts that some inequality
\begin{equation}
15~{\rm terms}~S_{RRRR} \geq 10~{\rm terms}~S_{RRR} + 6~{\rm terms}~S_{RRRRR}
\nonumber
\end{equation}
should be valid. The prediction that similar inequalities exist for every instance of (\ref{newpineq}) is a nontrivial corollary of our conjecture, which may aid ongoing and future efforts to describe the full holographic entropy cone. 

A simplifying observation is that inequalities~(\ref{newpineq}) need not be proven separately for each $N$. It suffices to prove them at the lowest $N$ that makes sense, $N = 2p-1$, where they take the simplified form~(\ref{specialodd}). For all higher $N'$, (\ref{newpineq}) follows from (\ref{specialodd}) by symmetrization over $S_{N'+1}$.  

\begin{acknowledgements}
\noindent
We thank Jan de Boer, Bowen Chen, Jingyuan Chen, Lampros Lamprou, Xiongfeng Ma, Xiaoliang Qi, Bogdan Stoica, Michael Walter, Zi-zhi Wang, and Daiming Zhang for useful discussions. We also thank Bogdan Stoica and Michael Walter for sharing with us their computer programs \cite{bogdanscontractor} and \cite{michaelscontractor}, and Daiming Zhang for contributing a proof of (\ref{upperbound}), which is reproduced in Part C of Supplemental Material.
\end{acknowledgements}

\newpage
\section*{Supplemental Material}

\subsection{Extreme vectors \mbox{of the conjectured cone of average entropies}}
\noindent
The extreme vectors are given by minimal cuts through flower graphs, which are shown in Figure~1 in the main text. 

\subsubsection{Computing the extreme rays}
\noindent
The computation of $S^p$ is as follows. First, $p$-partite regions fall into two classes: those where the stem of weight $w$ belongs to the $p$-partite region, and those where it belongs to the $(N+1-p)$-partite complement. The number of cases in the two classes is $\binom{N}{p-1}$ and $\binom{N}{p}$, respectively. 

In each case, we take the smaller of two cuts, which pass `in front of' or `behind' the core of the flower. (We decided against using the botanically correct term for the core, which is `ovary.') In the case where the stem is part of the $p$-partite region, the two cuts give $w+p-1$ and $N+1-p$. In the case where the stem is part of the $(N+1-p)$-partite complement, the two cuts give $p$ and $N-p + w$. We remind the reader that $1 \leq p \leq \myfloor{(N+1)/2}$. 

Collecting all possibilities, we have:
\begin{align}
\binom{N+1}{p} S^p & = 
\binom{N}{p-1} \min\!\big\{w+p-1, N+1-p\big\} \nonumber \\
& + \binom{N}{p} \, \min\{p, N-p+w\}
\label{detail}
\end{align}
Our range of interest is $1 \leq w \leq N$ because it contains all extreme rays of the cone of average entropies. For these values of $w$, expression~(\ref{detail}) simplifies to:
\begin{equation}
S^p = \frac{p}{N+1} \Big(N + \min\!\big\{w, N+2-2p\big\} \Big) 
\label{vectors}
\end{equation}
At fixed $w$, $\min\{ \ldots \} = w$ for $p \leq \tilde{p}$ and $(N+2-2p)$ for $p \geq \tilde{p}$, where:
\begin{equation}
\tilde{p} = (N+2-w)/2
\end{equation}
Substituting $w = N+2 - 2\tilde{p}$ in (\ref{vectors}), we find:
\begin{equation}
S^p = \frac{2}{N+1} \times \left\{\begin{array}{lp{1cm}l} p(N+1-\tilde{p}) && p \leq \tilde{p} \\
                                                                p(N+1-p) && p \geq \tilde{p} \end{array}\right.
\label{svectorapp}
\end{equation}
This is equation~(\ref{svector}) from the main text, with an altered normalization; see also Figure~4.

\subsubsection{Flowers with other stem weights $w$}
\noindent
We have argued that flower graphs describe stages of evaporation of black holes. But only flower graphs with weights, which range from $N$ to 1 or 2 in decrements of 2, are extreme rays of the holographic cone of average entropies. Here we briefly analyze other values of $w$.

First, consider $w$ in the interval $(1,N)$ or $(2,N)$, but not at one of the special values $w = N+2 - 2\tilde{p}$ where $1 \leq \tilde{p} \leq \myfloor{(N+1)/2}$ is integer. Average entropies of such flower graphs linearly interpolate between the extreme vectors attained at integer $\tilde{p}$. Therefore, by smoothly varying $w$ in this range, we can see the entropy vector traverse from one extreme ray to the next, eventually visiting all of them. Such a trajectory ends up drawing the shape of a simplex, which is a cross-section of the holographic cone of average entropies. We display such trajectories as dotted lines in Figures 2, 6, 7, 8.

When $w \geq N$, all minima in (\ref{detail}) are frozen at their $w$-independent values, which are attained at $w = N$. This range corresponds to black holes before Page time.

The remaining ranges are slightly more involved. Explicitly computing~(\ref{detail}) for $0 \leq w \leq 1$  gives:
\begin{align}
S^p & =  \frac{p(N+2w-1)}{N+1}  && {\rm if}~N~{\rm odd,}~p = \frac{N+1}{2} 
\label{exceptional} \\
S^p & = \frac{p(N+w)}{N+1} && {\rm otherwise}
\label{exceptotherwise}
\end{align}
First assume $N$ is even. Then the special circumstance~(\ref{exceptional}) never materializes and we find that all average entropies $S^p \propto p$. This is the entanglement pattern of an $N$-party perfect tensor (which for even $N$ is attained at $w = 2$), only rescaled by a constant. The same pattern holds for $1 \leq w \leq 2$.

When $N$ is odd, equations~(\ref{exceptional}-\ref{exceptotherwise}) satisfy the following identity:
\begin{equation}
S^p(w) = \frac{N+2w-1}{N+1}\,S^p(\tilde{w}) \quad {\rm with}~~\tilde{w} = \frac{2N-Nw+w}{N+2w-1}
\end{equation}
Since $0 \leq w \leq 1$ translates to $1 \leq \tilde{w} \leq 2N/(N-1)$, we see that the flowers with $0 \leq w \leq 1$ retrace the trajectory of $w \geq 1$, only with a changed normalization. 

\subsubsection{Verifying that extreme rays are extreme}
\label{appa2}
\noindent
Here we confirm that vectors $(\ref{svectorapp})$ live on the intersection of $\myfloor{(N+1)/2}-1$ hyperplanes $F_q = 0$. The quantities $F_q$ were defined in inequality~(\ref{newpineq}). We index hyperplanes with $q$ because $p$ is used for the average entropies $S^p$. The unique hyperplane, to which vector~(\ref{svectorapp}) does not belong, has $q = \tilde{p}$. 

Quantities $F_q$ involve only $S^{q-1}$ and $S^q$ and $S^{q+1}$. In a given vector~(\ref{svectorapp}), all three of them will fall in one range of $S^p$---linear or quadratic---unless $q = \tilde{p}$. But $S^p \propto p$ and $S^p \propto p (N+1-p)$ both satisfy $F_q = 0$. 

We should also verify that vector~(\ref{svectorapp}) lives on the `correct' side of the remaining hyperplane. Substituting (\ref{svector}) in (\ref{newpineq}) gives $F_{\tilde{p}} = 1$.

\subsection{Holographic cones of average entropies}
\subsubsection{$N \leq 4$}
\noindent
The holographic cones of average entropies for fewer than three named regions are exceptional. They live in a one-dimensional space, and are simply given by $S^1 \geq 0$. This can be seen as a special case of the inequality $2S^1 - S^2 \geq 0$, which bounds all higher cones of averages. The special circumstance occurs because $S^1 \equiv S^2$ at $N < 3$. (We have $S_A = S_O$ at $N=1$ and $S_{AB} = S_O$ at $N=2$.)

At $N=3$, the cone of averages lives in a two-dimensional space and is bounded by two inequalities:
\begin{align}
2 S^1 - S^2 & \geq 0 \\
- 4S^1 + 3S^2 & \geq 0
\end{align}
As explained in the main text, the first inequality is the symmetrization of subadditivity $S_A + S_B \geq S_{AB}$. At $N=3$, the second inequality is synonymous with the monogamy of mutual information 
\begin{equation}
S_{AB} + S_{BC} + S_{CA} - S_{A} - S_{B} - S_{C} - S_{ABC} \geq 0 \label{monogamyapp}
\end{equation}
because the latter is $S_4$-invariant. It is also inequality~(\ref{specialodd}) from the main text. The cone is shown in the left panel of Figure~6.

\begin{figure}[!t]
\label{n34fig}
\begin{center}
\includegraphics[width=0.45\columnwidth]{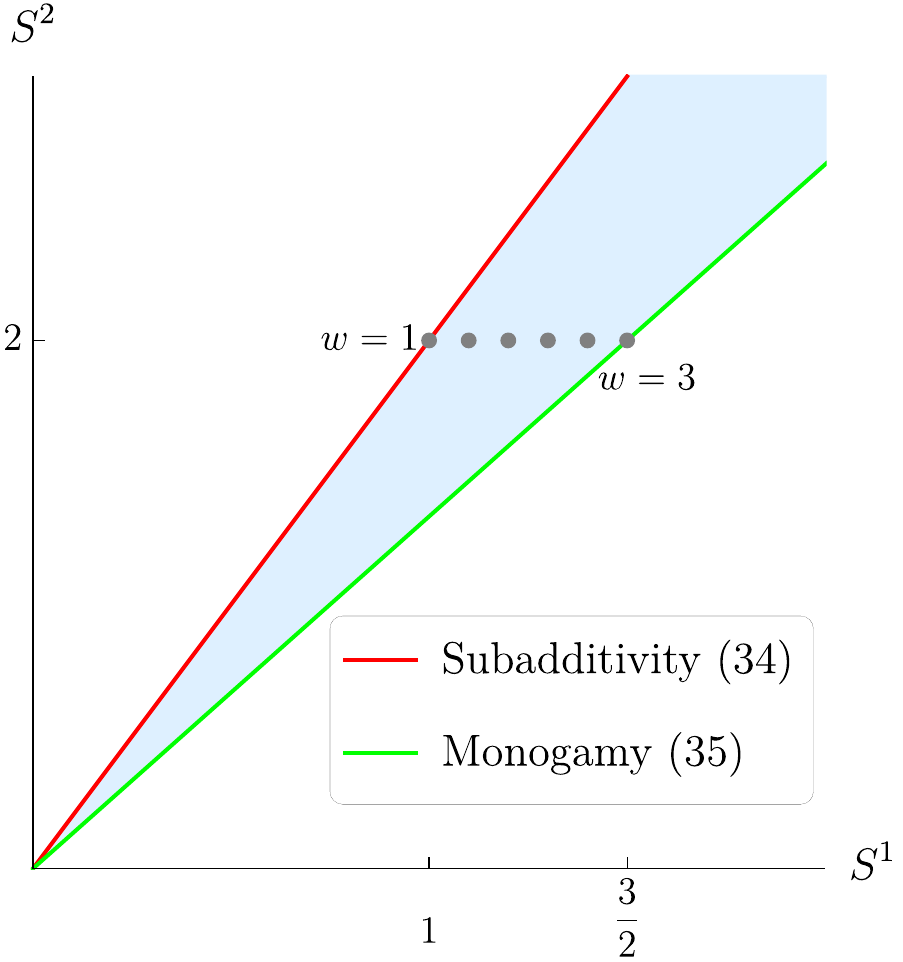}
\hspace*{0.05\columnwidth}
\includegraphics[width=0.45\columnwidth]{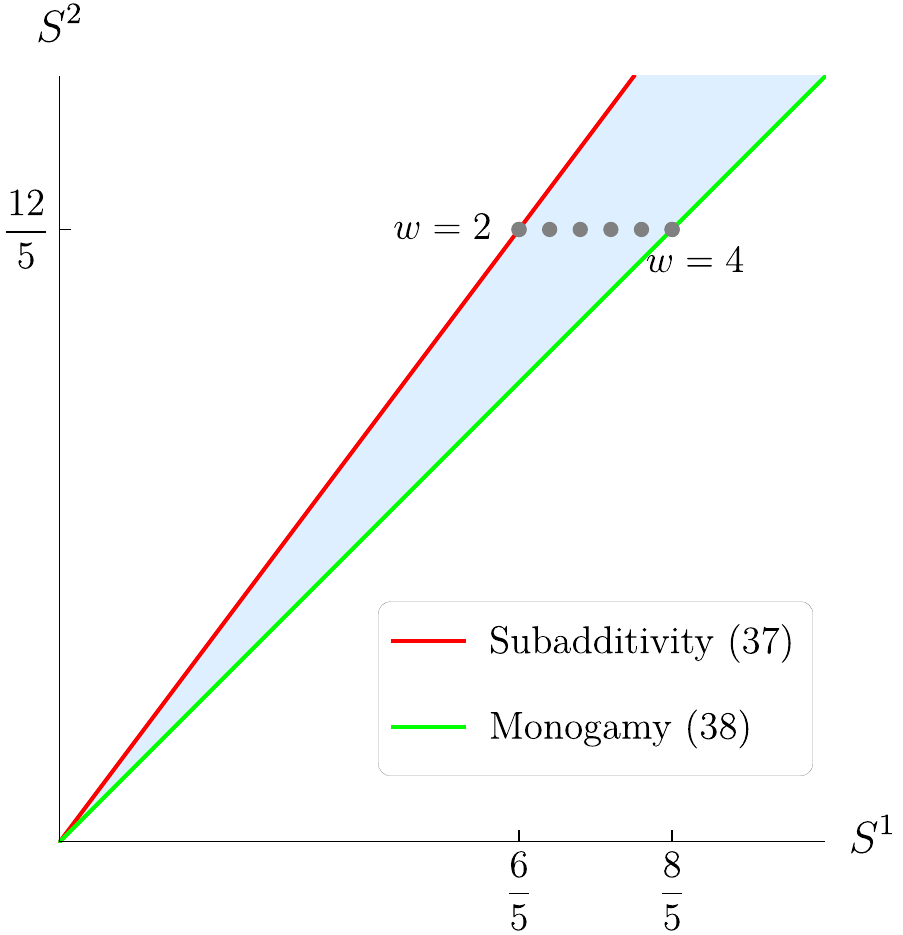}
\caption{The holographic cones of average entropies for $N=3$ (left) and $N=4$ (right) named regions. The dots mark average entropies obtained from cuts through flower graphs, which are computed in equation~(\ref{vectors}).}
\end{center}
\end{figure}

At $N=4$, the cone of averages lives in a two-dimensional space and is bounded by two inequalities:
\begin{align}
2 S^1 - S^2 & \geq 0 \\
- 3S^1 + 2S^2 & \geq 0
\end{align}
The second inequality symmetrizes~(\ref{monogamyapp}) with respect to $S_5$, so 2-partite and 3-partite terms contribute to $S^2$. It is also inequality~(\ref{specialeven}) from the main text. The cone is shown in the right panel of Figure~6. 

\subsubsection{$N = 5$}
\noindent
This cone of averages is shown in Figure~2 of the main text. It lives in a three-dimensional space and is bounded by three inequalities:
\begin{align}
2 S^1 - S^2 & \geq 0 \label{q1n5} \\
- 3S^1 + 3S^2 - S^3 & \geq 0 \label{q2n5} \\
- 9 S^2 + 8 S^3 & \geq 0 \label{q3n5}
\end{align}
The second inequality symmetrizes~(\ref{monogamyapp}) with respect to $S_6$, finally treating 1-partite, 2-partite, and 3-partite entropies as distinct terms. It is inequality~(\ref{newpineq}) from the main text with $p = 2$. The third inequality is (\ref{specialodd}) from the main text. It symmetrizes with respect to $S_6$ inequalities (\ref{e5}-\ref{e7}) of Reference~\cite{cuenca}:
\begin{align}
S_{ABC} & + S_{ABD} + S_{ABE} + S_{ACD} + S_{ACE} \nonumber \\
& + S_{ADE} + S_{BCE} + S_{BDE} + S_{CDE} \nonumber \\
& \qquad \qquad \qquad \geq \tag{E5} \label{e5} \\
S_{AB} & + S_{AC} + S_{AD} + S_{BE} + S_{CE} + S_{DE} \nonumber \\
& + S_{BCD} + S_{ABCE} + S_{ABDE} + S_{ACDE} \nonumber
\end{align}
and
\begin{align}
3S_{ABC} & + S_{ADE} + S_{ABE} + S_{ACD} + 3 S_{ACE} \nonumber \\
+\, 3S_{ABD}& + S_{BCD} + S_{BCE} + S_{BDE} + S_{CDE} \nonumber \\
& \qquad \qquad \qquad  \geq \tag{E6} \label{e6} \\
2S_{AB}\! +\! 2S_{AC} &\! +\! S_{AD}\! +\! S_{AE}\!+\! S_{BC}\! +\! 2S_{BD}\! +\! 2 S_{CE}\!+\! S_{DE} \nonumber \\
+\, 2 S_{ABCD} & + 2S_{ABCE} + S_{ABDE} + S_{ACDE} \nonumber
\end{align}
and
\begin{align}
2S_{ABC} + S_{ABD} & + S_{ABE} + S_{ACD} \nonumber \\
+\, S_{ADE} & + S_{BCE} + S_{BDE} \nonumber \\
&  \geq \tag{E7} \label{e7} \\
S_{AB} + S_{AC} + S_{AD} & + S_{BC} + S_{BE} + S_{DE} \nonumber \\
+ \, S_{ABCD} & + S_{ABCE} + S_{ABDE} \nonumber
\end{align}
The fact that all three of them share the same symmetrization suggests an underlying deeper structure. Our conjecture is an attempt to decipher that hint.  

In addition to subadditivity, monogamy~(\ref{monogamyapp}), and (\ref{e5}-\ref{e7}), three other classes ($S_6$ orbits) of inequalities bound the full $N=5$ holographic entropy cone \cite{cuenca}. One of them belongs to the infinite sequence of cyclic inequalities, which were first reported in \cite{hec}: 
\begin{align}
S_{ABC} + S_{BCD}\, + & \, S_{CDE} + S_{DEA} + S_{EAB} \nonumber \\ 
& \geq \label{cyclicineq} \\
S_{AB} + S_{BC} + S_{CD} \, + & \, S_{DE} + S_{EA} + S_{ABCDE} \nonumber
\end{align}
This symmetrizes to:
\begin{equation}
5S^3 - 5S^2 - S^1 \geq 0
\label{cyclicsymm}
\end{equation}
Another one is monogamy (\ref{monogamy}) with substitutions $B \to BD$ and $C \to CE$:
\begin{equation}
S_{ABD} + S_{BCDE} + S_{ACE} - S_{A} - S_{BD} - S_{CE} - S_{ABCDE} \geq 0 \label{monogamycomp}
\end{equation}
The last one is listed in \cite{cuenca} as (E8) and reads:
\begin{align}
S_{AD} + S_{BC} + S_{ABE} + S_{ACE} & + S_{ADE} + S_{BDE} + S_{CDE} \nonumber \\
&  \geq \tag{E8} \label{e8} \\ 
S_{A} + S_{B}&  + S_{C} + S_{D} \nonumber \\
+ \, S_{AE} + S_{DE} + S_{BCE} & + S_{ABDE} + S_{ACDE} \nonumber
\end{align}
Both (\ref{monogamycomp}) and (\ref{e8}) symmetrize to:
\begin{equation}
-2 S^1 - S^2 + 2 S^3 \geq 0
\label{e8symm}
\end{equation}
at $N=5$ named regions because 2-partite and 4-partite entropies are complementary. (Their symmetrizations at higher $N$ differ.) Inequalities~(\ref{cyclicsymm}) and (\ref{e8symm}) are strictly weaker than (\ref{q1n5}-\ref{q3n5}).  

\begin{figure}[!t]
\label{n6fig}
\begin{center}
\includegraphics[width=0.45\columnwidth]{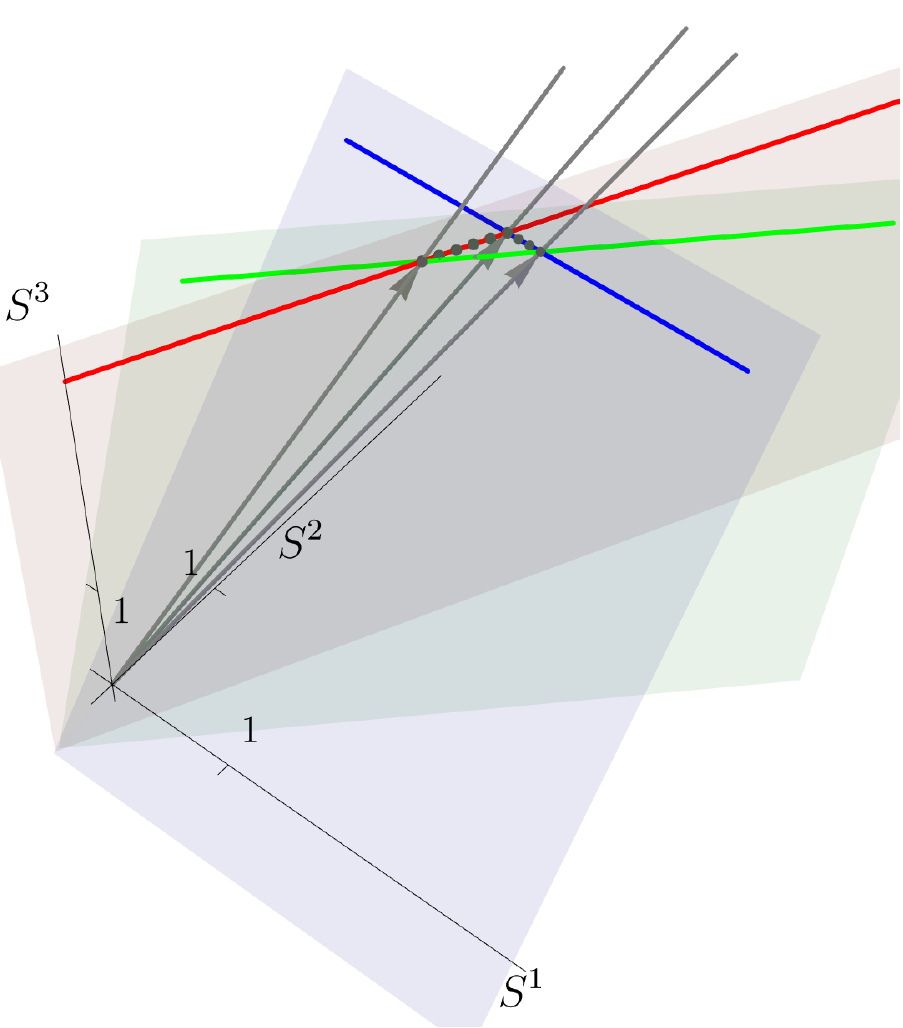}
\hspace*{0.05\columnwidth}
\includegraphics[width=0.45\columnwidth]{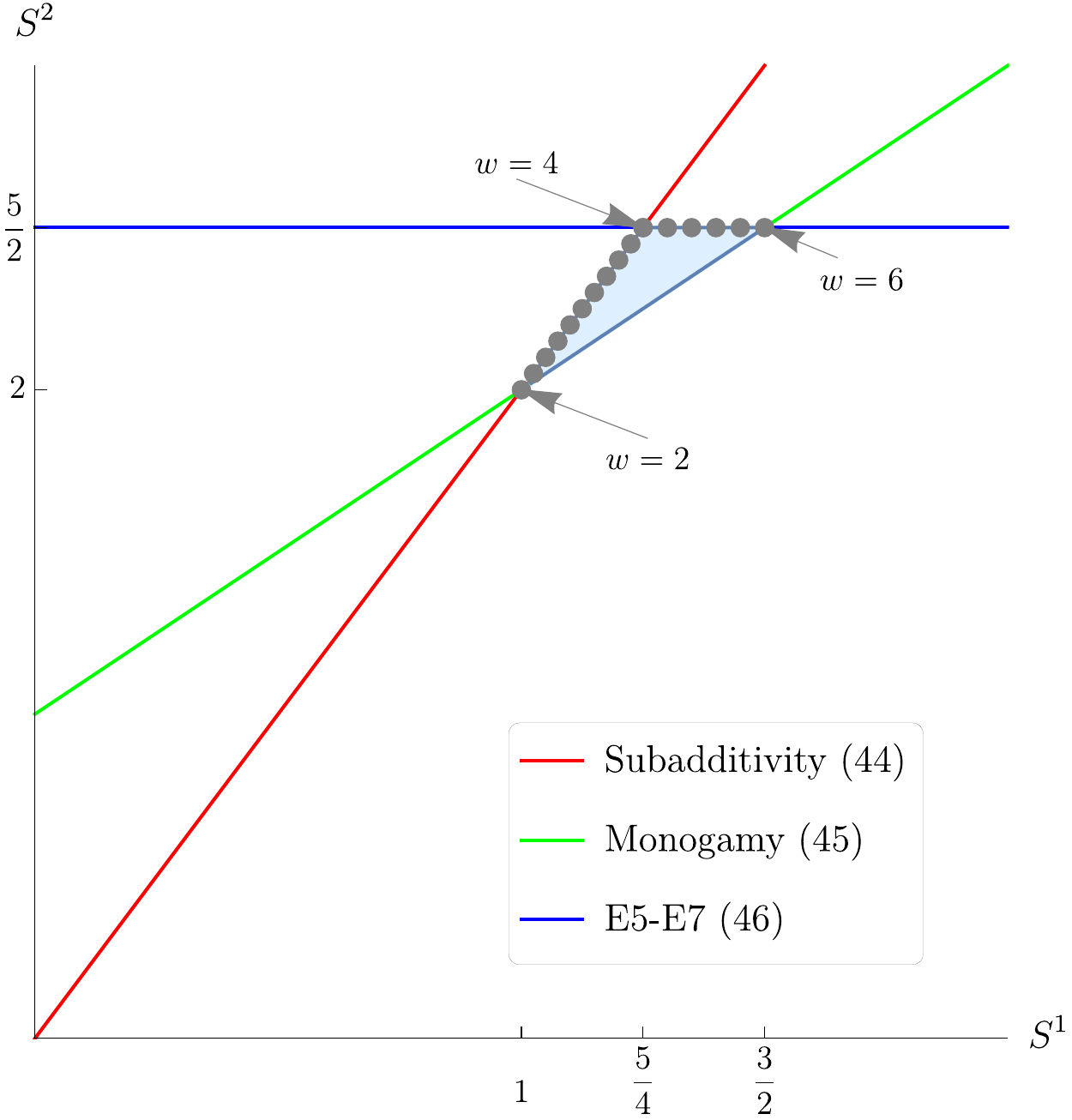}
\caption{The holographic cone of average entropies for $N=6$ named regions: the three-dimensional view of $(S^1, S^2, S^3)$-space and the two-dimensional cross-section $S^3 = 3$. The cone is bound by three planes, where inequalities~(\ref{q1n6}-\ref{q3n6}) are saturated. Entropies obtained from flower graphs~(\ref{vectors}), displayed as grey dots, have been rescaled by 7/8.}
\end{center}
\end{figure}

\subsubsection{$N = 6$}
\noindent
The holographic cone of average entropies for $N=6$ named regions is shown in Figure~7. It lives in a three-dimensional space and is bounded by three inequalities:
\begin{align}
2 S^1 - S^2 & \geq 0 \label{q1n6} \\
- 3S^1 + 3S^2 - S^3 & \geq 0 \label{q2n6} \\
- 6 S^{2} + 5 S^{3} & \geq 0 \label{q3n6}
\end{align}
We already know that the second one is a symmetrization of monogamy~(\ref{monogamyapp}), and that it is inequality~(\ref{newpineq}) from the main text with $p=2$. The third one symmetrizes (\ref{e5}-\ref{e7}) under the permutation group $S_7$, which treats 3-partite and 4-partite entropies as equivalent. That inequality is (\ref{specialeven}) from the main text.

At present, the complete $N=6$ holographic entropy cone is not fully known. But at the level of average $p$-partite entropies, we know there are no new inequalities because the entire cone defined by (\ref{q1n6}-\ref{q3n6}) can be realized with cuts through weighted graphs. 

\subsubsection{The conjectured cone at $N = 7$ named regions}
\noindent
The cone is shown in Figure~8. This is the first case where we can only conjecture the holographic cone of average entropies rather than prove it. The cone lives in a four-dimensional space, and the conjecture posits that it is demarcated by four inequalities. 

We first present known facts about the $N=7$ cone of averages and then turn to conjectural assertions. The following inequalities are known:
\begin{align}
2 S^1 - S^2 & \geq 0 \label{q1n7} \\
- 3S^1 + 3S^2 - S^3 & \geq 0 \label{q2n7} \\
- 6 S^{2} + 8 S^3 - 3 S^{4} & \geq 0 \label{q3n7} \\
- 3 S^2 -4 S^3 + 6 S^4 & \geq 0 \label{q4n7} \\
-S^1 - 7 S^3 + 7 S^4 & \geq 0 \label{q5n7}
\end{align}
The first two symmetrize subadditivity and monogamy. The third one is the $S_8$ symmetrization of inequalities~(\ref{e5}-\ref{e7}). The fourth one symmetrizes over $S_8$ inequality (\ref{e7}) after the substitution $A \to AF$. The fifth one symmetrizes the $N = 7$ cyclic inequality of \cite{hec}. These five form the tightest set of currently known inequalities bounding $N=7$ average entropies. We have verified that other known inequalities produce strictly weaker conditions. 

\begin{figure}[!t]
\label{n7fig}
\begin{center}
\includegraphics[width=0.45\columnwidth]{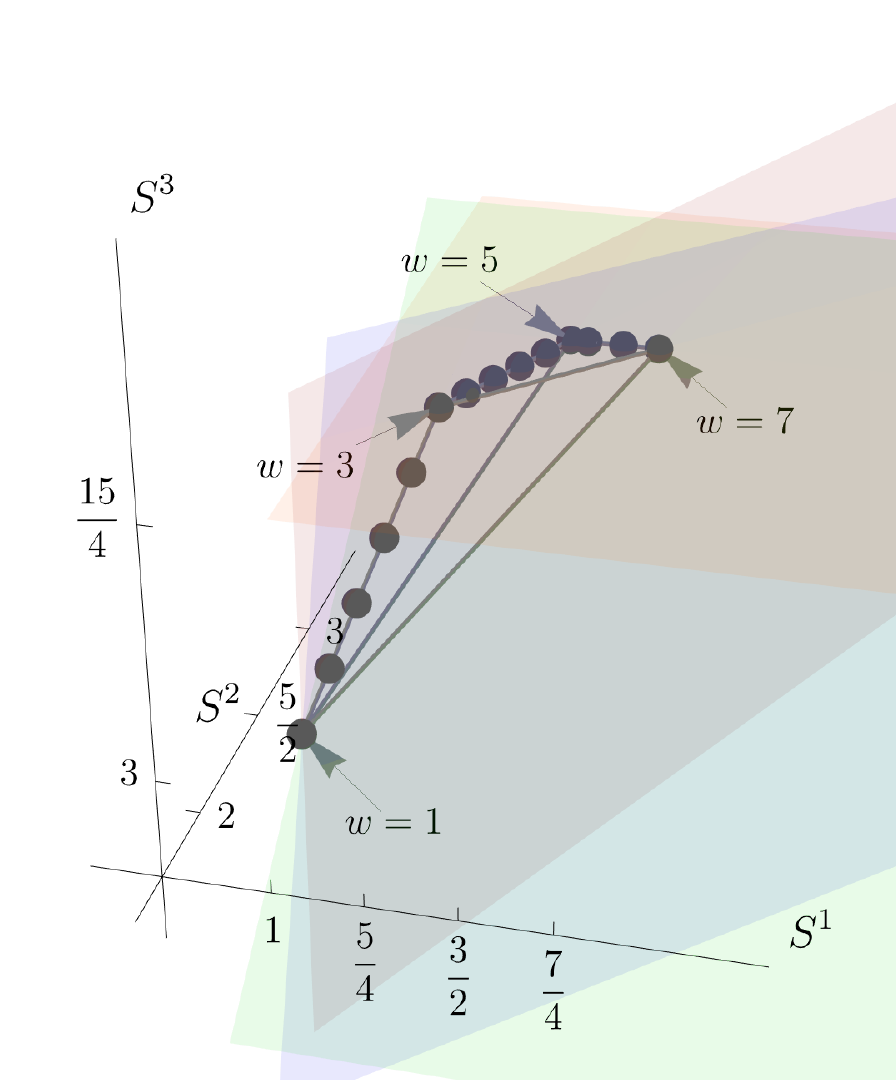}
\hspace*{0.05\columnwidth}
\includegraphics[width=0.45\columnwidth]{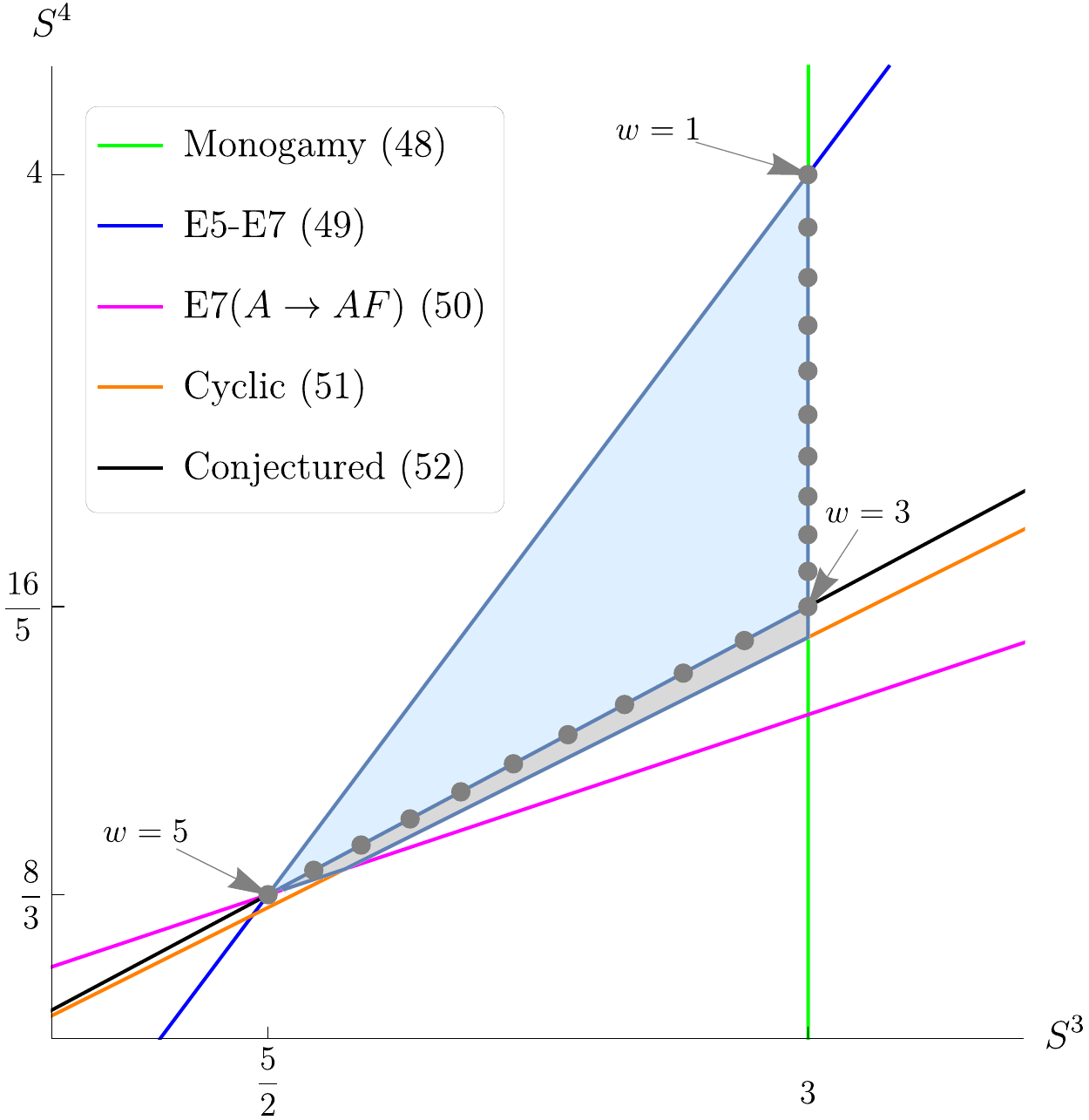}
\caption{The conjectured holographic cone of average entropies for $N=7$ named regions. Because the cone lives in four dimensions, we show two cross-sectional views: a three-dimensional view of $(S^1, S^2, S^3)$-space at $S^4 = 4$ and a two-dimensional view of $(S^3, S^4)$-space at $S^1 = 1$ and $S^2 = 2$. In the left panel, the average entropies obtained from flower graphs (equation~\ref{vectors}) clearly mark the tetrahedral cross-section of the conjectured cone. In the right panel, we mark known and conjectured inequalities, which bound the $N=7$ average entropies. Subadditivity~(\ref{q1n7}) is not indicated explicitly because it is saturated on the plane $(S^1, S^2) = (1,2)$. The thin, dark-filled quadrilateral near the bottom of the figure is allowed by known inequalities (\ref{q4n7}) and (\ref{q5n7}), but excluded by inequality~(\ref{conjn7}) and our conjecture.}
\end{center}
\end{figure}

Despite numerous efforts, we have been unable to find any weighted graphs whose cuts violate the following inequality:
\begin{equation}
-16 S^3 + 15 S^4 \geq 0
\label{conjn7}
\end{equation}
This is inequality~(\ref{specialodd}) from the main text. If this inequality holds, it makes~(\ref{q4n7}) and (\ref{q5n7}) redundant. In qualitative terms, the sliver of entropy space allowed by (\ref{q4n7}) and (\ref{q5n7}) but excluded by (\ref{conjn7}) is small; see Figure~8.

\subsection{Upper bound (\ref{upperbound}) on $S^{p+1} / S^p$}
\label{proofpb}
\noindent
We thank Daiming Zhang for contributing the following proof of inequality:
\begin{equation}
\frac{S^{p+1}}{S^{p}} \leq \frac{p+1}{p} \tag{\ref{upperbound}}
\end{equation}
The proof only uses strong subadditivity \cite{ssaref}, which establishes that (\ref{upperbound}) holds generally in quantum mechanics.

We start with a definition and a lemma. Let us apply subadditivity $S_X + S_Y - S_{XY} \geq 0$ to an individual (1-partite) $X$ and a $p$-partite $Y$. After symmetrization, this defines a nonnegative quantity $I_p$:
\begin{equation}
I_p \equiv S^1 + S^{p} - S^{p+1} \geq 0
\label{defip}
\end{equation}
Now consider strong subadditivity $S_{XZ} + S_{YZ} - S_{XYZ} - S_Z \geq 0$. (Note that subadditivity is a special case of strong subadditivity with $Z = \emptyset$.) Choosing a $(p-1)$-partite $Z$ and 1-partite $X$ and $Y$ and symmetrizing, we find:
\begin{equation}
2S^p - S^{p+1} - S^{p-1} = I_p - I_{p-1} \geq 0
\label{ssalemma}
\end{equation}
From the definition of $I_p$ we have:
\begin{align}
S^p = S^1 + S^{p-1} - I_{p-1} & = 2S^1 + S^{p-2} - I_{p-1} - I_{p-2} \nonumber \\
\ldots & = p S^1 - \sum_{p' = 1}^{p-1} I_{p'}
\end{align}
By lemma~(\ref{ssalemma}), this implies:
\begin{equation}
S^p \geq pS^1 - (p-1) I_p
\end{equation}

Using these, we obtain:
\begin{align}
\frac{S^{p+1}}{S^p} & = 1 + \frac{S^{p+1} - S^p}{S^p} = 1 + \frac{S^1 - I_p}{S^p} \nonumber \\
& \leq 1 + \frac{S^1 - I_p}{pS^1 - (p-1) I_p} \equiv f(I_p / S^1)
\label{intermediate}
\end{align}
where
\begin{equation}
f(x) = \frac{p + 1 - px}{p - (p-1) x}
\end{equation}
We have $f(0) = (p+1)/p$ so inequality (\ref{upperbound}) may be saturated by $x = I_p / S^1 = 0$. Because of (\ref{ssalemma}), the inequality in (\ref{intermediate}) has an equal sign at $I_p = 0$ and saturation of (\ref{upperbound}) is indeed achieved. Finally, the derivative is $f'(x) = - (p+x-px)^{-2}$ so $f(x) < (p+1)/p$ for all positive values of $x = I_p / S^1$. Therefore, for all $I_p > 0$ inequality~(\ref{upperbound}) is strict.

We found that (\ref{upperbound}) is saturated if and only if $I_p = 0$. But then (\ref{defip}) and (\ref{ssalemma}) sandwich all other $I_{p' < p}$ into vanishing. Therefore, saturation of (\ref{upperbound}) at $p$ implies saturation of the same bound for all $p' < p$, as claimed in the main text.

\newpage


\begin{thebibliography}{99}

\bibitem{adscft}
J.~M.~Maldacena,
``The large $N$ limit of superconformal field theories and supergravity,''
Adv. Theor. Math. Phys. \textbf{2} (1998), 231-252
[arXiv:hep-th/9711200 [hep-th]].

\bibitem{marksessay}
M.~Van Raamsdonk,
``Building up spacetime with quantum entanglement,''
Gen. Rel. Grav. \textbf{42} (2010), 2323-2329
[arXiv:1005.3035 [hep-th]].

\bibitem{erepr}
J.~Maldacena and L.~Susskind,
``Cool horizons for entangled black holes,''
Fortsch. Phys. \textbf{61} (2013), 781-811
[arXiv:1306.0533 [hep-th]].

\bibitem{hec}
N.~Bao, S.~Nezami, H.~Ooguri, B.~Stoica, J.~Sully and M.~Walter,
``The holographic entropy cone,''
JHEP \textbf{09} (2015), 130
[arXiv:1505.07839 [hep-th]].

\bibitem{rt1}
S.~Ryu and T.~Takayanagi,
``Holographic derivation of entanglement entropy from AdS/CFT,''
Phys. Rev. Lett. \textbf{96} (2006), 181602
[arXiv:hep-th/0603001 [hep-th]].

\bibitem{rt2}
S.~Ryu and T.~Takayanagi,
``Aspects of holographic entanglement entropy,''
JHEP \textbf{08} (2006), 045
[arXiv:hep-th/0605073 [hep-th]].

\bibitem{hrt}
V.~E.~Hubeny, M.~Rangamani and T.~Takayanagi,
``A Covariant holographic entanglement entropy proposal,''
JHEP \textbf{07} (2007), 062
[arXiv:0705.0016 [hep-th]].

\bibitem{bh1}
J.~D.~Bekenstein,
``Black holes and entropy,''
Phys. Rev. D \textbf{7}, 2333-2346 (1973)

\bibitem{bh2}
S.~W.~Hawking,
``Black hole explosions,''
Nature \textbf{248}, 30-31 (1974)

\bibitem{mmiref}
P.~Hayden, M.~Headrick and A.~Maloney,
``Holographic mutual information is monogamous,''
Phys. Rev. D \textbf{87} (2013) no.4, 046003
[arXiv:1107.2940 [hep-th]].

\bibitem{cuenca}
S.~Hern\'andez Cuenca,
``Holographic entropy cone for five regions,''
Phys. Rev. D \textbf{100} (2019) no.2, 2
[arXiv:1903.09148 [hep-th]].

\bibitem{arrangement}
V.~E.~Hubeny, M.~Rangamani and M.~Rota,
``The holographic entropy arrangement,''
Fortsch. Phys. \textbf{67} (2019) no.4, 1900011
[arXiv:1812.08133 [hep-th]].

\bibitem{kbasis}
T.~He, M.~Headrick and V.~E.~Hubeny,
``Holographic entropy relations repackaged,''
JHEP \textbf{10} (2019), 118
[arXiv:1905.06985 [hep-th]].

\bibitem{superbalance}
T.~He, V.~E.~Hubeny and M.~Rangamani,
``Superbalance of holographic entropy inequalities,''
JHEP \textbf{07} (2020), 245
[arXiv:2002.04558 [hep-th]].

\bibitem{mmipt}
S.~X.~Cui, P.~Hayden, T.~He, M.~Headrick, B.~Stoica and M.~Walter,
``Bit threads and holographic monogamy,''
Commun. Math. Phys. \textbf{376} (2019) no.1, 609-648
[arXiv:1808.05234 [hep-th]].

\bibitem{diffent}
V.~Balasubramanian, B.~D.~Chowdhury, B.~Czech, J.~de Boer and M.~P.~Heller,
``Bulk curves from boundary data in holography,''
Phys. Rev. D \textbf{89} (2014) no.8, 086004
[arXiv:1310.4204 [hep-th]].

\bibitem{saref}
H.~Araki and E.~H.~Lieb, ``Entropy inequalities,'' 
Communications in Mathematical Physics {\bf 18}, 160 (1970).

\bibitem{sm}
In Part A of Supplemental Material we write down the vectors of average entropies for flower graphs at general $N$ and verify that those vectors are extreme rays of the cone of average entropies if and only if inequalities (\ref{newpineq}) are its facets. In Part B we give detailed forms of holographic cones of average entropies for $N \leq 7$ regions, marking for every facet its parent inequality from the full holographic entropy cone. In Part C, we share Daiming Zhang's proof of inequality~(\ref{upperbound}) from strong subadditivity.

\bibitem{ssaref}
E.~H.~Lieb and M.~B.~Ruskai,
``Proof of the strong subadditivity of quantum-mechanical entropy,''
J. Math. Phys. \textbf{14}, 1938-1941 (1973).

\bibitem{happy}
F.~Pastawski, B.~Yoshida, D.~Harlow and J.~Preskill,
``Holographic quantum error-correcting codes: Toy models for the bulk/boundary correspondence,''
JHEP \textbf{06}, 149 (2015)
[arXiv:1503.06237 [hep-th]].

\bibitem{random}
P.~Hayden, S.~Nezami, X.~L.~Qi, N.~Thomas, M.~Walter and Z.~Yang,
``Holographic duality from random tensor networks,''
JHEP \textbf{11}, 009 (2016)
[arXiv:1601.01694 [hep-th]].

\bibitem{eprref}
A.~Einstein, B.~Podolsky and N.~Rosen, 
``Can quantum-mechanical description of physical reality be considered complete?,''
Phys. Rev. \textbf{47} (1935), 777-780.

\bibitem{bitthreads}
M.~Freedman and M.~Headrick,
``Bit threads and holographic entanglement,''
Commun. Math. Phys. \textbf{352} (2017) no.1, 407-438
[arXiv:1604.00354 [hep-th]].

\bibitem{scrambling}
Y.~Sekino and L.~Susskind,
``Fast scramblers,''
JHEP \textbf{10} (2008), 065
[arXiv:0808.2096 [hep-th]].

\bibitem{page1}
D.~N.~Page,
``Information in black hole radiation,''
Phys. Rev. Lett. \textbf{71} (1993), 3743-3746
[arXiv:hep-th/9306083 [hep-th]].

\bibitem{page2}
D.~N.~Page,
``Time dependence of Hawking radiation entropy,''
JCAP \textbf{09} (2013), 028
[arXiv:1301.4995 [hep-th]].

\bibitem{penington}
G.~Penington,
``Entanglement wedge reconstruction and the information paradox,''
JHEP \textbf{09} (2020), 002
[arXiv:1905.08255 [hep-th]].

\bibitem{princetonucsb}
A.~Almheiri, N.~Engelhardt, D.~Marolf and H.~Maxfield,
``The entropy of bulk quantum fields and the entanglement wedge of an evaporating black hole,''
JHEP \textbf{12} (2019), 063
[arXiv:1905.08762 [hep-th]].

\bibitem{princeton}
A.~Almheiri, R.~Mahajan, J.~Maldacena and Y.~Zhao,
``The Page curve of Hawking radiation from semiclassical geometry,''
JHEP \textbf{03} (2020), 149
[arXiv:1908.10996 [hep-th]].

\bibitem{thankxlq}
We thank Xiaoliang Qi for this insightful observation. 

\bibitem{pagethm}
D.~N.~Page,
``Average entropy of a subsystem,''
Phys. Rev. Lett. \textbf{71} (1993), 1291-1294
[arXiv:gr-qc/9305007 [gr-qc]].

\bibitem{octopi}
C.~Akers, N.~Engelhardt and D.~Harlow,
``Simple holographic models of black hole evaporation,''
JHEP \textbf{08} (2020), 032
[arXiv:1910.00972 [hep-th]].

\bibitem{firewalls}
A.~Almheiri, D.~Marolf, J.~Polchinski and J.~Sully,
``Black holes: Complementarity or firewalls?,''
JHEP \textbf{02} (2013), 062
[arXiv:1207.3123 [hep-th]].

\bibitem{bogdanscontractor}
B.~Stoica, an implementation of the greedy algorithm for finding contraction maps on Mathematica\textsuperscript{\textregistered}. 

\bibitem{michaelscontractor}
M.~Walter, ``The holographic contractor,'' program in \CC, Copyright 2015.

%
%
%
%
%
%
%



\end{thebibliography}
\end{document}